\documentclass[aps,preprint,superscriptaddress]{revtex4}
\usepackage{graphicx}		% Einbinden von Grafiken

\begin{document}

\title{Morphogen Transport in Epithelia}

\author{T. Bollenbach} 
\affiliation{Max-Planck-Institute for the Physics of Complex Systems, N\"othnitzer
Strasse~38, 01187 Dresden, Germany} 
\author{K. Kruse} 
\affiliation{Max-Planck-Institute for the Physics of Complex Systems, N\"othnitzer
Strasse~38, 01187 Dresden, Germany} 
\author{P. Pantazis} 
\affiliation{Max-Planck-Institute of Molecular Cell Biology and Genetics, Pfotenhauer
Strasse~108, 01307 Dresden, Germany} 
\author{M. Gonz\'alez-Gait\'an} 
\affiliation{Max-Planck-Institute of Molecular Cell Biology and Genetics, Pfotenhauer
Strasse~108, 01307 Dresden, Germany} 
\author{F. J\"ulicher}
\affiliation{Max-Planck-Institute for the Physics of Complex Systems, N\"othnitzer
Strasse~38, 01187 Dresden, Germany}

\date{\today}

\begin{abstract}
We present a general theoretical framework to discuss mechanisms of
morphogen transport and gradient formation in a cell layer.
Trafficking events on the cellular scale lead to transport on larger
scales. We discuss in particular the case of transcytosis where
morphogens undergo repeated rounds of internalization into cells and
recycling. Based on a description on the cellular scale, we derive
effective nonlinear transport equations in one and two dimensions
which are valid on larger scales. We derive analytic expressions for
the concentration dependence of the effective diffusion coefficient
and the effective degradation rate. We discuss the effects of a
directional bias on morphogen transport and those of the coupling of
the morphogen and receptor kinetics. Furthermore, we discuss general
properties of cellular transport processes such as the robustness of
gradients and relate our results to recent experiments on the
morphogen Decapentaplegic (Dpp) that acts in the fruit fly Drosophila.
\end{abstract}

\maketitle 

%\tableofcontents

\section{Introduction}

Morphogens are signaling molecules which are secreted from cells in a
restricted source region and provide signals to cells located at a
distance from this source.  They play a key role for the determination
of cell fates in animal development \cite{taba04}.  While the term
``morphogen'' was coined by Turing in his seminal work on pattern
formation in reaction-diffusion systems \cite{turi52}, the modern
paradigm of morphogen action was introduced by Wolpert in 1969
\cite{wolp69}.  According to this paradigm, morphogens spread from the
source region into the adjacent target tissue where they are partly
degraded. The combination of the localized production of morphogens,
transport, and degradation leads to the formation of a non-equilibrium
steady state in which the morphogen concentration decreases with
increasing distance from the morphogen source. This concentration
profile is called ``morphogen gradient'' and contains positional
information about the distance from the morphogen source. Cells in the
tissue detect the local morphogen concentration via receptor molecules
that are present on their surface and respond by expressing a set of
target genes in a manner that depends on the detected morphogen
concentration.  In this way, the morphogen gradient can generate a
pattern of differentiated cells in the target tissue.  In the last two
decades, the existence of morphogen gradients has been supported by
considerable experimental evidence.  Prominent examples of signaling
molecules that function as morphogens are Bicoid which acts in the
embryo of the fruit fly Drosophila \cite{drie88a,drie88b},
Decapentaplegic (Dpp) which acts in the Drosophila wing disk
\cite{entc00,tele00}, and Activin which acts in the embryo of the frog
Xenopus \cite{gurd94}.

The mechanisms by which morphogens are transported and gradients are
formed are so far not well understood.  A difficulty in the study of
morphogen kinetics is the fact that morphogen transport in a tissue is
coupled to cellular trafficking processes.  It is influenced for
example by ligand-receptor binding, the endocytosis of ligand-receptor
pairs, and the kinetics of receptor numbers.  For a long time, it was
taken for granted that morphogens move by diffusion in the
extracellular space surrounding the cells \cite{cric70}. In a few
cases, there is experimental evidence for this: the morphogen Activin
in Xenopus is an example \cite{mcdo97}. However, for one of the best
studied model systems, the morphogen Dpp in the Drosophila wing disk,
experiments have called diffusive transport into question and
suggested an important role of cell surface molecules in this process
\cite{entc00,bele04}. Consequently, other transport mechanisms than
extracellular diffusion have been suggested
\cite{entc00,bele04,rami99}. Firstly, Dpp could be transported by
transcytosis. Here, transport is achieved via repeated rounds of
morphogen binding to cell surface receptors, internalization into the
cell and subsequent externalization, and release of the ligand from
the receptor at a different position on the cell surface
\cite{entc00,krus04}.  Secondly, Dpp might move by passive diffusion
on the cell surface. Here, a certain type of large molecules (called
Heparan Sulfate Proteoglycans, HSPGs) which are located on the cell
surface could function as a ``carrier'' for the morphogens
\cite{bele04}. Furthermore, morphogen transport could occur in
cytonemes which are long membrane tubes that connect the morphogen
source cells to cells in the target tissue \cite{rami99}.  Due to the
complexity of the problem, a combination of theoretical descriptions
of morphogen gradient formation and systematic experiments is needed
to identify the dominant morphogen transport mechanism
\cite{land02,krus04}.

 During animal development, the precision of the positions of
differentiating cells in the tissue and the times at which cells
differentiate is typically high \cite{houc02,greg05}.  This indicates
that robust mechanisms that are insensitive to changes of
environmental conditions and to intrinsic fluctuations have evolved to
control cell differentiation.  Clearly, such robustness could be
achieved if morphogen gradients are themselves robust.  Evidence for
the robustness of morphogen gradients was found in recent experiments
\cite{mori96,elda02,elda03}.  This robustness must originate in the
mechanisms by which morphogens are transported and degraded. The
robustness and precision of morphogen gradients
\cite{elda02,elda03,boll05,greg05,howa05} or a possible role
of anomalous diffusion in morphogen transport \cite{horn05} can only
be understood using a combination of theoretical and experimental
efforts.

In this article, we provide a detailed description of morphogen
gradient formation by different mechanisms and provide full
derivations of the morphogen transport equations.  Some of our results
have been recently published in \cite{boll05}. Our description
captures several processes that are supported by experimental data for
the morphogen Dpp in the Drosophila wing disk. These experimental
findings will be briefly summarized in section
\ref{sec:wing_disk}. While we focus on the wing disk of the fruit fly
here, these processes are very likely to play an important role for
morphogen gradient formation in other animals.  Starting from a
description of cellular trafficking processes, we derive in section
\ref{sec:tc1d} effective transport equations on larger scales. We
investigate key properties of gradient formation by these mechanisms
in section \ref{sec:eff_trans} and study the effects of a directional
bias for transport resulting from cellular polarity.  Finally, we
extend our approach to higher dimensions in section \ref{sec:2d_desc}
and discuss morphogen transport in two dimensional epithelia.

\section{Morphogen gradient formation and cellular trafficking processes }\label{sec:wing_disk}

The larva of the fruit fly Drosophila contains precursors of the
organs of the adult animal. The precursor of the fly wing is a flat
pouch that consists of two cell layers that are connected at the edges
and is called wing disk (see Fig.~1 in \cite{krus04}). The thicker one
of these cell layers is formed by columnar epithelial cells and
includes the so-called wing primordium.  In the following, we consider
this two dimensional cell layer \cite{krus04,land02}.  Dpp is produced
and secreted in a specific source region which is a narrow stripe with
a width of about 7 cell diameters that is located at the center of
this layer.  Cells outside of this source do not produce Dpp but
posses receptors located at their cell surface to detect its
presence. Dpp spreads from the source region into the adjacent target
tissue on both sides of the source region.  In the whole tissue, Dpp
molecules are degraded. As a consequence of the localized source and
degradation, a graded morphogen profile is built up. This formation of
the morphogen gradient can be directly observed in experiments by
using a Dpp that is labeled with green fluorescent protein (GFP-Dpp)
\cite{entc00,tele00}. In steady state, the Dpp gradient extends over
$50\mu {\rm m}$ into the target tissue. This corresponds to about 20
cell diameters.

Several cellular processes are relevant during the formation of this
morphogen gradient.  Morphogens are ligands which bind to specific
receptor molecules. Ligand-receptor pairs are internalized into the
cell via endocytosis. Internalized ligands are either degraded or they
can be recycled to the cell surface via exocytosis. Finally,
receptor-ligand pairs can diffuse on the cell surface \cite{kers98}
and free ligands can diffuse in the extracellular space surrounding
the cells.

Furthermore, cells in the wing disk produce, degrade, externalize, and
internalize  receptor molecules. In general, the
production rate of these receptors is affected by the local morphogen
concentration. For example, a high Dpp concentration leads to a
reduced production  rate of the Dpp receptor Thick-veins (Tkv) in the wing
disk \cite{tele00}. This behavior is called ``receptor
down-regulation''. The opposite behavior can also occur: high
concentrations of Hedgehog (Hh), another morphogen acting in the wing
disk, lead to an increased production of its receptor Patched (Ptc)
\cite{chen96}. This phenomenon is called ``receptor up-regulation''.

Recent experiments have revealed the interplay of Dpp gradient
formation and cellular trafficking.  Firstly, endocytosis has been
blocked in the whole wing disk except for the Dpp source region or,
alternatively, in smaller patches of cells (clones) in the tissue
\cite{entc00,krus04}. This has been achieved using mutant flies in
which endocytosis can be blocked at an elevated temperature of
$34^\circ {\rm C}$ due to the temperature-sensitive mutation shibire.
Five hours upon blocking endocytosis in the whole target tissue,
GFP-Dpp fluorescence was almost undetectable in the target tissue
while its gradient extended visibly over more than 20 cell diameters
into this tissue before the endocytic block. When endocytosis was only
blocked in patches of cells near the Dpp source region, a pronounced
transient depletion of the GFP-Dpp concentration behind these clones
was observed. Both experiments indicate a role of endocytosis in Dpp
transport. This suggests that instead of simply diffusing in the
extracellular space, Dpp is transported via the cell interior in
repeated rounds of endocytosis and exocytosis. This transport
mechanism is called transcytosis \cite{entc00}.  Indeed, a theoretical
description in which morphogen transport is solely based on
extracellular diffusion while interactions of the morphogen with its
receptors are taken into account was found to be inconsistent with the
experimental observations \cite{krus04}. 

In a different set of experiments, the role of large cell surface
molecules (HSPGs) in morphogen transport was investigated
\cite{bele04}.  A depletion of extracellular Dpp was observed behind
mutant clones of cells which lack HSPGs. This indicates a role of
HSPGs in Dpp transport. Indeed, it has been suggested that HSPGs
facilitate the diffusion of morphogens on the cell surface
\cite{bele04}.

\section{Morphogen transport in one dimension} \label{sec:tc1d} 

We introduce a discrete description of morphogen transport by
transcytosis and passive extracellular diffusion and derive transport
equations for these processes. The theoretical description developed
here is generally applicable to biological systems in which molecules
are transported by extracellular diffusion and transcytosis.  To
stress this generality, we will mostly refer to the transported
molecules as ``ligands'' instead of ``morphogens''.

\subsection{Ligand kinetics in a chain of cells}\label{sec:1d_disc}

In one space dimension, we describe the ligand kinetics in a chain of
cells, see Fig.~\ref{fig:ptc_scheme}. We denote the distance between
the centers of two neighboring cells by $a$ and the width of the gap
between two cells by $b$.  In this chain, $n$ indexes the cells, see
Fig.~\ref{fig:ptc_scheme}.  The number of free extracellular ligands
between cells $n$ and $n+1$ is denoted $L_n$.  The numbers of
intracellular free and ligand-bound receptors are denoted $R_n^{\rm
(i)}$ and $S_n^{\rm (i)}$, respectively.  $R_n^{\rm (l)}$ and
$R_n^{\rm (r)}$ denote the concentrations of free receptors on the
left and right cell surfaces, respectively.  $S_n^{\rm (l)}$ and
$S_n^{\rm (r)}$ denote the ligand-bound receptors accordingly.  The
kinetics of the ligand and receptor numbers are given by
\begin{eqnarray}
\frac{d}{dt} L_n & = & k_{\rm off} (S_n^{\rm (r)} + S_{ n+1}^{\rm (l)}) - k_{\rm on}
(R_n^{\rm (r)} + R_{ n+1}^{\rm (l)}) L_n  + \frac{D_0}{a^2}
(L_{n+1}+L_{n-1}-2 L_n) - e_{\rm deg} L_n\cr
\frac{d}{dt} R_n^{\rm (r)} & = & \frac{f_{\rm syn}}{2} + k_{\rm off} S_n^{\rm (r)}
- k_{\rm on} R_n^{\rm (r)}L_{ n} - f_{\rm int} R_{ n}^{\rm (r)} +
\frac{f_{\rm  ext}}{2} R_n^{\rm (i)} \cr
\frac{d}{dt} S_n^{\rm (r)} & = & -k_{\rm off} S_n^{\rm (r)} + k_{\rm on} R_n^{\rm
  (r)}L_{ n} - b_{\rm int} S_{ n}^{\rm (r)} +  \frac{b_{\rm ext}}{2} S_n^{\rm
  (i)} \cr 
\frac{d}{dt} R_n^{\rm (l)} & = &  \frac{f_{\rm syn}}{2} + k_{\rm off} S_n^{\rm
  (l)} - k_{\rm on} R_n^{\rm (l)}L_{ n-1} - f_{\rm int} R_{ n}^{\rm (l)} +
\frac{f_{\rm ext}}{2} R_n^{\rm (i)} \cr 
\frac{d}{dt} S_n^{\rm (l)} & = & -k_{\rm off} S_n^{\rm (l)} + k_{\rm on}
R_n^{\rm 
  (l)}L_{ n-1} - b_{\rm int} S_{ n}^{\rm (l)} + \frac{b_{\rm ext}}{2} S_n^{\rm
  (i)} \cr 
\frac{d}{dt} R_n^{\rm (i)} & = & - f_{\rm ext} R_n^{\rm (i)} + f_{\rm int} (R_{ n}^{\rm
  (l)} + R_n^{\rm (r)})  - f_{\rm deg} R_n^{\rm (i)}\cr 
\frac{d}{dt} S_n^{\rm (i)} & = & - b_{\rm ext} S_n^{\rm (i)} + b_{\rm int} (S_{ n}^{\rm
  (l)} + S_n^{\rm (r)}) - b_{\rm deg} S_n^{\rm (i)}. \label{eq:ptc_full_disc}
\end{eqnarray}
Here, the binding and un-binding of ligands to and from receptors is
characterized by rates $k_{\rm on}$ and $k_{\rm off}$.  The
internalization and externalization of receptor-ligand complexes is
captured by the rates $b_{\rm int}$ and $b_{\rm ext}$.  Free ligands
can hop directly from one gap between the cells to the adjacent ones
at a rate $2D_0/a^2$. This describes their free diffusion in the
extracellular space around the cells with diffusion coefficient $D_0$.
The degradation of ligands in the extracellular space occurs with rate
$e_{\rm deg}$ and that of ligands bound to receptors inside the cell
with rate $b_{\rm deg}$.  Furthermore, free receptors are internalized
and externalized with rates $f_{\rm int}$ and $f_{\rm ext}$
respectively.  Internalized free receptors are degraded with rate
$f_{\rm deg}$.  In addition, each cell produces receptors with a rate
$f_{\rm syn}$.

Newly produced receptors appear on the cell surface \cite{lauf93}.
The rate of receptor synthesis $f_{\rm syn}$ in
Eq. (\ref{eq:ptc_full_disc}) depends on $R_n^{\rm (l)}+R_n^{\rm (r)}$ and
$S_n^{\rm (l)}+S_n^{\rm (r)}$:
\begin{equation}
f_{\rm syn} = f_{\rm syn}^0 \left(1-\frac{R_n^{\rm (l)}+R_n^{\rm
    (r)}+\psi\left(S_n^{\rm (l)}+S_n^{\rm (r)}\right)}{R_{\rm max}}\right),
\label{eq:rec_syn_disc}
\end{equation}
where $f_{\rm syn}^0$ is a basal rate of receptor synthesis and
$R_{\rm max}$ is the saturation value of the surface receptor
concentration at which the production of new receptors stops.  The
dimensionless parameter $\psi$ couples receptor synthesis to the
concentration of morphogens. The two cases of receptor up- and
down-regulation \cite{lauf93} are captured by $\psi<1$ and $\psi>1$,
respectively.  

The description (\ref{eq:ptc_full_disc}) is valid in the bulk of the
system.  We still have to specify the kinetics at the boundaries of
the chain.  To describe the effects of a ligand source located at
$n=0$ in Eq. (\ref{eq:ptc_full_disc}), we modify the equation for the free
ligand:
\begin{equation}
\frac{d}{dt} L_0 = k_{\rm off} S_{1}^{\rm (l)} - k_{\rm on} R_{
1}^{\rm (l)} L_0 + \frac{D_0}{a^2} (L_{1}- L_0) - e_{\rm deg} L_0 +
\nu,
\label{eq:source_bound_disc}
\end{equation}
where $\nu$ is the rate at which ligands from the source enter the
system.  At the position $n=N$ where the lattice ends, we impose
\begin{equation}
\frac{d}{dt} L_N = k_{\rm off} S_{N}^{\rm (r)} - k_{\rm on} R_{N}^{\rm
(r)} L_N + \frac{D_0}{a^2} (L_{N-1}- L_N) - e_{\rm deg} L_N,
\label{eq:wall_bound_disc}
\end{equation}
which describes a zero flux boundary condition at this edge of the
lattice. This boundary condition can be motivated by the geometry of
the Drosophila wing disk \cite{krus04}. However, if $N$ is
sufficiently large the ligand number at $N$ is small and the choice of
boundary condition has only a small influence on the ligand profile.
A sequence of morphogen and receptor profiles at different times that
were obtained by numerical solution of Eq. (\ref{eq:ptc_full_disc}) are
presented in Fig.~\ref{fig:comp_rec_dyn} together with the steady
state profiles that represent the morphogen gradient.

\subsection{Effective transport equations on larger scales}\label{sec:1d_theor_tc}\label{sec:1d_cont}
We derive effective continuum transport equations for ligand transport
starting from Eq. (\ref{eq:ptc_full_disc}). We introduce the
concentrations $ l(t,x)=L_n(t)/a,\ r^{\rm (l)}(t,x)=R_n^{\rm
(l)}(t)/a,\ r^{\rm (r)}(t,x)=R_n^{\rm (r)}(t)/a,\ r_i(t,x)=R_n^{\rm
(i)}(t)/a, \ s^{\rm (l)}(t,x)=S_n^{\rm (l)}(t)/a,\ s^{\rm
(r)}(t,x)=S_n^{\rm (r)}(t)/a,\ \textnormal{and }\ s_i(t,x)=S_n^{\rm
(i)}(t)/a $ where $x=na$.  We derive continuum equations for the
kinetics of these densities starting from Eq. (\ref{eq:ptc_full_disc}) by
locally expanding the densities in a power series with respect to $x$,
for example $L_{n+1}/a=l(x+a)=l(x)+a \partial_x l(x)+ a^2 \partial_x^2 l(x)/2$.
It is further useful to introduce the new variables $r_\pm(t,x) =
r^{\rm (l)}(t,x)\pm r^{ (r)}(t,x)$ and $s_\pm(t,x)=s^{\rm (l)}(t,x)\pm
s^{\rm (r)}(t,x)$ so that $r_+$ and $s_+$ measure the total free and
ligand bound surface receptor concentrations per cell and $r_-$ and
$s_-$ the polarization of these concentrations on the cell surface,
respectively.

In situations where the length $\xi_D$ over which the steady state
gradient decays is large compared to the cell diameter $a$, a
separation of time scales occurs in the system which makes the
adiabatic elimination of rapid variables possible. Indeed, if $\tau_a$
is the relaxation time of the kinetics within one cell, the slow
relaxation of the gradient occurs on a time scale $\tau_{\xi_D}=\tau_a
(\xi_D/a)^2 \gg \tau_a$. We thus use the approximation that all local
kinetics relaxes instantaneously.  At each position $x$, this yields
the relations
\begin{eqnarray} \label{eq:ad_ap_f_1d}
l & = & \frac{k_{\rm off}s_+}{k_{\rm on}a r_+}  \cr
s_i & = & \frac{b_{\rm int} s_+}{b_{\rm ext}}  \cr
s_- & = & \frac{k_{\rm on} a l r_- -  \frac{ab k_{\rm on}}{2} r_+ \partial_x l 
+  \frac{(a-b) b_{\rm ext}}{2} \partial_x s_i}{b_{\rm int} + k_{\rm off}}  \cr
r_i & = & \frac{f_{\rm int} r_+}{f_{\rm ext}}  \cr
r_- & = & \frac{k_{\rm off} s_- +  \frac{ab k_{\rm on}}{2} r_+ \partial_x l 
+  \frac{(a-b)f_{\rm ext}}{2} \partial_x r_i}{f_{\rm int} + k_{\rm on} a l}.  \label{eq:full_le}
\end{eqnarray}

Using these expressions, we can adiabatically eliminate five of the
seven variables $l,s_i,s_+,s_-,r_i,r_+,r_-$ and obtain only two
coupled equations for the remaining slow variables which are the total
ligand density $\lambda(x,t)=l(x,t)+s_i(x,t)+s_+(x,t)$ and the total
receptor density $\rho(x,t)=r_i(x,t)+r_+(x,t)+s_i(x,t)+s_+(x,t)$:
\begin{eqnarray}
\partial_t \lambda & = & \partial_x\left(D_\lambda(\lambda, \rho)\partial_x\lambda +D_\rho(\lambda, \rho)\partial_x\rho\right)
- k_\lambda(\lambda, \rho) \lambda, \label{eq:lambda_ptc_eq_full}\\
\partial_t \rho & = & \nu_{\rm syn}(\lambda,\rho) - k_\rho(\lambda, \rho) \rho. \label{eq:rho_ptc_eq_full}
\end{eqnarray}
The other densities $l,s_i,s_+,s_-,r_i,r_+,r_-$ can be calculated from
$\lambda$, $\rho$, and their first spatial derivatives via
Eq. (\ref{eq:full_le}).  The derivation of Eqs. (\ref{eq:full_le}),
(\ref{eq:lambda_ptc_eq_full}), and (\ref{eq:rho_ptc_eq_full}) is
discussed in appendix \ref{sec:deri}. In addition to equations
(\ref{eq:lambda_ptc_eq_full}) and (\ref{eq:rho_ptc_eq_full}), this
procedure provides us with explicit expressions for the effective
diffusion coefficient $D_\lambda$, the effective degradation rate
$k_\lambda$, the receptor degradation rate $k_\rho$ as well as the
transport coefficient $D_\rho$ which describes ligand transport
induced by gradients of the receptor concentration, see
Eq. (\ref{eq:dk_expl_1d}).  In Fig.~\ref{fig:dlambda}~A,B, the
coefficients $D_\lambda$, $D_\rho$ and $k_\lambda$ are displayed as a
function of the ligand concentration $\lambda$ for a typical choice of
parameters.  The effective receptor production rate $\nu_{\rm
syn}(\lambda,\rho)=f_{\rm syn}/a$ is a function of the ligand and
receptor concentrations.  The functional form of $\nu_{\rm syn}$ that
corresponds to the receptor production rate (\ref{eq:rec_syn_disc}) in
the discrete description is
\begin{equation}
\nu_{\rm syn}(\lambda,\rho) = \frac{f_{\rm syn}^0}{a}
 \left(1-\frac{r_+(\lambda,\rho) + \psi s_+(\lambda,\rho)}{r_{\rm
 max}}\right),
\label{eq:nu_1d}
\end{equation}
with $r_{\rm max}=R_{\rm max}/a$. Ligand transport is associated with
the current $j=-(D_\lambda(\lambda, \rho)\partial_x\lambda
+D_\rho(\lambda, \rho)\partial_x\rho)$. The current $j_0$ at $x=0$ is
related to the secretion rate $\nu$ in
Eq. (\ref{eq:source_bound_disc}). In one dimension  $j_0=\nu$.

In Fig.~\ref{fig:comp_rec_dyn} we show time-dependent receptor and
ligand profiles which are solutions to
Eqs. (\ref{eq:lambda_ptc_eq_full}) and
(\ref{eq:rho_ptc_eq_full}). These solutions are in agreement with the
corresponding solutions to the discrete description
(\ref{eq:ptc_full_disc}) which validates the adiabatic approximation
made in the derivation of Eqs. (\ref{eq:lambda_ptc_eq_full}) and
(\ref{eq:rho_ptc_eq_full}).

\subsection{Effects of a directional bias}\label{sec:bias}

If cells possess a polarity, transcytosis can have a bias and lead to
directed transport. Here, we include a directional bias in our
description of ligand transport. Experiments indicate that the
transport of the morphogen Dpp in the Drosophila wing disk is
non-directional on macroscopic length scales \cite{entc00}. However,
epithelia with cell polarity could in principle exhibit directional
transcytosis.

We therefore generalize the discrete description
(\ref{eq:ptc_full_disc}) by allowing receptor-bound ligand molecules
to be preferentially externalized on either the left or the right cell
surface. To this end, we introduce a dimensionless parameter $\beta$
which measures this bias: for $\beta=-1/2$, all receptor-bound ligands
are externalized on the left surface, for $\beta=1/2$ on the right
surface, and for $\beta=0$, we recover the unbiased description, see
appendix \ref{sec:ptc_bias} for details.

In presence of such a bias, the transport equations
(\ref{eq:lambda_ptc_eq_full}) and (\ref{eq:rho_ptc_eq_full}) generalize to
\begin{eqnarray}
\partial_t \lambda & = & \partial_x(D_\lambda(\lambda,
  \rho)\partial_x\lambda +D_\rho(\lambda, \rho)\partial_x\rho -
  V_\beta(\lambda,\rho) \lambda)
- k_\lambda(\lambda, \rho) \lambda, \cr
\partial_t \rho & = & \nu_{\rm syn}(\lambda,\rho) - k_\rho(\lambda, \rho) \rho \label{eq:ptc_eq_full_bias},
\end{eqnarray}
where $V_\beta(\lambda,\rho)$ is a concentration-dependent effective
drift velocity, see Eq. (\ref{eq:vbeta_expl}).  The other coefficients in
Eq. (\ref{eq:ptc_eq_full_bias}) remain the same as in the case without
directional bias.  In Fig.~\ref{fig:bias}~A, we show $V_\beta$ as a
function of $\lambda$ for $\beta=0.1$, i.e. a bias that leads to
preferential transport to the right.  Typical ligand and receptor
profiles that are generated in presence of a bias are shown in
Fig.~\ref{fig:bias}~B,C.

\subsection{Constant surface receptor concentration}\label{sec:const_r}

We now  discuss the simple case where the
total surface receptor number $R$ is constant everywhere.  This
approximation is useful because it still captures most
important features of morphogen transport by transcytosis
\cite{boll05}.

In this case the receptor kinetics in Eq. (\ref{eq:ptc_full_disc}) becomes
obsolete.  The equations describing ligand transport in this simple
case are
\begin{eqnarray}
\frac{d}{dt} L_n & = & k_{\rm off} (S_n^{\rm (r)} + S_{ n+1}^{\rm (l)}) - k_{\rm on}
(R - S_n^{\rm (r)} - S_{ n+1}^{\rm (l)} ) L_n  + \frac{D_0}{a^2} (L_{n+1}+L_{n-1}-2 L_n) - e_{\rm deg} L_n\cr
\frac{d}{dt} S_n^{\rm (r)} & = & -k_{\rm off} S_n^{\rm (r)} + k_{\rm on}
\Bigl(\frac{R}{2}-S_n^{\rm (r)}\Bigr) L_{ n} - b_{\rm int} S_{ n}^{\rm (r)} + \frac{1}{2} b_{\rm ext} S_n^{\rm (i)} \cr
\frac{d}{dt} S_n^{\rm (l)} & = & -k_{\rm off} S_n^{\rm (l)} + k_{\rm on}
\Bigl(\frac{R}{2}-S_n^{\rm (l)}\Bigr) L_{ n-1} - b_{\rm int} S_{ n}^{\rm
  (l)} + \frac{1}{2} b_{\rm ext} S_n^{\rm (i)} \cr
\frac{d}{dt} S_n^{\rm (i)} & = & - b_{\rm ext} S_n^{\rm (i)} + b_{\rm
  int} (S_{ n}^{\rm (l)} + S_n^{\rm (r)}) - b_{\rm deg} S_n^{\rm (i)} \label{eq:disc_const_r}.
\end{eqnarray}
Boundary conditions analogous to Eqs. (\ref{eq:source_bound_disc}) and
 (\ref{eq:wall_bound_disc}) are imposed at $n=0$ and $n=N$.

If the surface receptor concentration is constant, the continuum limit
after adiabatic elimination of fast variables is described by
\begin{equation}
\label{eq:dldt_rconst}
\partial_t\lambda=\partial_x\left(D(\lambda)\partial_x\lambda\right)-k(\lambda)\lambda\quad,
\end{equation}
where the effective diffusion coefficient $D(\lambda)$ in the absence
of extracellular diffusion (for $D_0=0$) and the effective degradation
rate $k(\lambda)$ are given by
\begin{eqnarray}
D(\lambda) & = & \frac{a^2{b_{\rm ext}}{b_{\rm int}}{k_{\rm off}}a
   k_{\rm on} r C_{-}(\lambda) }{4 A(\lambda) ( 2a k_{\rm on} r k_{\rm
   off}({b_{\rm ext}}+ {b_{\rm int}}) + {b_{\rm int}}C_{-}(\lambda) )
   } \cr 
k(\lambda) & = & \frac{C_{+}(\lambda)}{a k_{\rm on}\lambda}
   \left( \frac{b_{\rm deg}{b_{\rm int}}}{2 {b_{\rm ext}}\left(
   {b_{\rm ext}} + {b_{\rm int}} \right) } + \frac{{e_{\rm
   deg}}{k_{\rm off}}}{ C_{-}(\lambda)} \right).
\label{eq:dk_constr_expl}
\end{eqnarray}
In these expressions, $r=R/a$ and
\begin{eqnarray}
A(\lambda) & = & \left[-4 b_{\rm ext} (b_{\rm ext}+b_{\rm int})a^2 k_{\rm on}^2 r\lambda+(
  b_{\rm int} a k_{\rm on} r + b_{\rm ext} B_{+}(\lambda))^2\right]^{1/2} \cr 
  B_{\pm}(\lambda) & = & {k_{\rm off}} + a k_{\rm on}(\lambda\pm r )\cr
C_{\pm}(\lambda) & = & b_{\rm int}a k_{\rm on} r\mp A(\lambda)\pm
  b_{\rm ext}B_{\pm}(\lambda). \nonumber
%\label{eq:dk_constr_abc}
\end{eqnarray}
These coefficients $D(\lambda)$ and $k(\lambda)$ are shown as a
function of $\lambda$ in Fig.~\ref{fig:dlambda}~C. Their non-linear
dependence on the total ligand concentration $\lambda$ is very similar
to that of the coefficients $D_\lambda(\lambda,\rho)$ and
$k_\lambda(\lambda,\rho)$ in Eq. (\ref{eq:lambda_ptc_eq_full}), see
Fig.~\ref{fig:dlambda}~A.

\vspace{1cm} We can describe transport by diffusion of receptor-bound
ligands in the cell membrane by the same methods. We discuss this
mechanism in appendix \ref{sec:glyp} where we consider the case where
endocytosis and recycling are unimportant. Furthermore, we have so far
discussed the case where extracellular diffusion is weak ($D_0$
small). The opposite case in which extracellular diffusion dominates
ligand transport is discussed in appendix \ref{sec:theor_dbts}.

\section{Properties of transport by transcytosis}\label{sec:eff_trans}

\subsection{Nonlinear diffusion and degradation}\label{sec:prop_tc}
Several key features of transcytosis follow directly from the general
shape of the transport equations (\ref{eq:lambda_ptc_eq_full}) and
(\ref{eq:rho_ptc_eq_full}) and from the dependence of the transport
and degradation coefficients on the ligand and receptor
concentrations.  Firstly, the presence of the term
$D_\rho(\lambda,\rho) \partial_x \rho$ in
Eq. (\ref{eq:lambda_ptc_eq_full}) shows that gradients of the receptor
concentration induce a ligand current that is directed towards higher
receptor concentrations since $D_\rho<0$.  This contribution to the
current comes up, because the ligand affinity for a region increases
with the receptor concentration in that region. For small ligand
concentrations $D_\rho \sim \lambda$ which ensures that the
corresponding current vanishes.  Furthermore, $D_\lambda$ and
$k_\lambda$ in Eq. (\ref{eq:lambda_ptc_eq_full}) approach finite values
for small $\lambda$. In this limit, Eq. (\ref{eq:lambda_ptc_eq_full})
consequently becomes a linear diffusion equation with
degradation. This implies that on large length scales and for small
ligand concentrations, transcytosis is indistinguishable from passive
diffusion.

In the opposite limit of large ligand concentrations $\lambda$,
$D_\lambda$ and $k_\lambda$ exhibit the asymptotic behavior
$$D_\lambda\simeq D_0 + c_1(\rho) \lambda^{-2}$$ and $k_\lambda\simeq
e_{\rm deg} + c_2(\rho) \lambda^{-1}$.  Here $D_0$ is the
extracellular diffusion coefficient which is approached in the limit
of large $\lambda$, and we have defined $c_1(\rho)= a b_{\rm ext}
f_{\rm int} k_{\rm off} \rho/4 k_{\rm on} (b_{\rm ext}+b_{\rm int})$
and $c_2(\rho)= b_{\rm deg} b_{\rm int} \rho/(b_{\rm ext}+b_{\rm
int})$.  The transport coefficient
$$D_\rho\simeq -D_0 -c_1(\rho) \lambda^{-2}$$ in this limit.
Interestingly, both $D_\lambda$ and $|D_\rho|$ approach the value
$D_0$. This implies that transport is dominated by extracellular
diffusion for large ligand concentrations $\lambda$.  This behavior
results from the fact that most receptors are occupied and can
consequently not participate in ligand transport by transcytosis.  A
maximum of $D_\lambda$ can occur for intermediate values of $\lambda$
as long as $D_0$ is smaller than a critical value, see
Fig.~\ref{fig:dlambda}~A. Similarly, there can be a minimum of
$D_\rho$ as a function of $\lambda$, see Fig.~\ref{fig:dlambda}~B. The
observation that ligand transport is most efficient at a specific
ligand concentration $\lambda$ is an interesting consequence of the
nonlinearities of the transport process.

In the special case $D_0=0$, $D_{\rho}$ as well as $D_\lambda$ vanish
in the absence of receptors, i.e. for $\rho=0$, or if either binding
or un-binding of ligands from the receptor, internalization or
externalization of occupied or free receptors is suppressed, i.e. if
either one of the rates $k_{\rm on}$, $k_{\rm off}$, $b_{\rm int}$,
$b_{\rm ext}$, $f_{\rm int}$, or $f_{\rm ext}$ vanishes. This reflects
that in the absence of extracellular diffusion, transport is generated
by repeated internalization and externalization of ligand-bound
receptors as well as ligands binding to and un-binding from surface
receptors.  In the limit of fast internalization or fast un-binding,
the ligands are confined to the cell interior or the extracellular
space, respectively, and transport is consequently hampered. Indeed
for $D_0=0$, $D_\lambda\to0$ if $b_{\rm int}\to\infty$ or if $k_{\rm
off}\to\infty$. We discuss other limits of our description in appendix
\ref{sec:simp_lim}.

Similarly, if transcytosis has a directional bias, the effective drift
velocity $V_\beta$ vanishes if either $\rho, k_{\rm on}, k_{\rm
off}, b_{\rm int}, b_{\rm ext}, f_{\rm int},$ or $f_{\rm ext}$ is zero
and also for $b_{\rm int}\to\infty$ or $k_{\rm
off}\to\infty$. Moreover, the drift velocity $V_\beta$ is
independent of $D_0$.  For small $\lambda$, $V_\beta$ adopts a finite
value.  $|V_\beta|$ is a monotonically decreasing function of
$\lambda$ and, in the limit of large $\lambda$, $V_\beta$ vanishes as
$V_\beta \simeq c_3(\rho) \lambda^{-2}$, with $c_3(\rho)=\beta b_{\rm
ext} f_{\rm int} k_{\rm off} \rho/[(b_{\rm ext}+b_{\rm int})k_{\rm
on}]$, see Fig.~\ref{fig:bias}~A. The fact that $V_\beta$ vanishes
asymptotically for large $\lambda$ again reflects that transport is
mediated by receptors which are only present in limited numbers.

\subsection{Steady state concentration profile} \label{sec:an_stst}

We now calculate the steady state ligand profile formed in the half
space $x\geq 0$ in the presence of a source which is located at $x<0$.
In the steady state, Eq. (\ref{eq:rho_ptc_eq_full}) with the condition
$\partial_t\rho=0$ yields a relation $\rho(x)=\rho_{\rm
s}(\lambda(x))$ between the total receptor concentration $\rho$ and
the total ligand concentration $\lambda$ at position $x$. This
combined with Eq. (\ref{eq:lambda_ptc_eq_full}) leads to the steady state
equation for the ligand profile
\begin{equation}
  \partial_x (D_{\rm s}(\lambda) \partial_x\lambda) - k_{\rm s}(\lambda) \lambda = 0, 
\label{eq:stst_eqn_full}  
\end{equation}
with the effective diffusion coefficient in the steady state $D_{\rm
s}(\lambda)=D_\lambda(\lambda,\rho_{\rm s}(\lambda)) +
D_\rho(\lambda,\rho_{\rm s}(\lambda)) d\rho_{\rm s}(\lambda)/d\lambda$
and the effective degradation rate $k_{\rm s}(\lambda)=
k_\lambda(\lambda,\rho_{\rm s}(\lambda))$.  The steady state relation
$\rho_{\rm s}(\lambda)$ is a monotonic function of $\lambda$ and
converges to finite values $\rho_\infty$ for $\lambda\to\infty$ and
$\rho_0$ for $\lambda\to 0$.  This reflects that each cell only
contains a limited number of receptors and is never completely devoid
of receptors.  It implies that $d\rho_{\rm s}/d\lambda=0$ for large
$\lambda$, so that in this limit $D_{\rm s}(\lambda)\simeq
D_\lambda(\lambda,\rho_\infty)$.

The steady state ligand profile $\lambda(x)$ described by
 Eq. (\ref{eq:stst_eqn_full}) can be determined exactly.  We rewrite
 Eq. (\ref{eq:stst_eqn_full}) as
\begin{eqnarray}
\partial_x j_{\rm s} & = & -k_{\rm s}(\lambda) \lambda \cr
\partial_x \lambda & = & - \frac{j_{\rm s}(\lambda)}{D_{\rm s}(\lambda)},
\nonumber
\end{eqnarray}
where the steady state current $j_{\rm s}$ is a function of
$\lambda(x)$ only. This implies
\begin{eqnarray}
\frac{dj_{\rm s}(\lambda)}{d\lambda} & = & \frac{k_{\rm s}(\lambda)\lambda D_{\rm
    s}(\lambda)}{j_{\rm s}(\lambda)} \cr
\frac{dx(\lambda)}{d\lambda} & = & -\frac{D_{\rm s}(\lambda)}{j_{\rm s}(\lambda)},\label{eq:stst_int2}
\end{eqnarray}
where $x(\lambda)$ is the inverse function of the steady state ligand
profile $\lambda(x)$. Using Eq. (\ref{eq:stst_int2}), we find the steady
state solution
\begin{equation}
x=-\int_{\lambda(0)}^{\lambda(x)}d\lambda^{\prime} 
D_{\rm s}(\lambda^{\prime})/j_{\rm s}(\lambda^\prime) \quad, \label{eq:stst}
\end{equation}
where the steady state current is
\begin{equation}
j_{\rm s}(\lambda) = \left(2\int_0^{\lambda}d\lambda^\prime\
k_{\rm s}(\lambda^\prime) D_{\rm s}(\lambda^\prime) \lambda^\prime\right)^{1/2}.  \label{eq:stst_current}
\end{equation}
In the steady state, the total ligand concentration decreases
monotonically with increasing distance to the source.  For small
$\lambda$, the ligand profile decays as $\lambda \sim \exp(-x/\xi)$
with $\xi=\sqrt{D_{\rm s}(0)/k_{\rm s}(0)}$.  For large ligand
concentrations $\lambda\gg \lambda_T$ and in the absence of free
diffusion (i.e. for $D_0=0$), the current $j_{\rm s}$ behaves asymptotically
as
$$j_{\rm s}^2(\lambda)\simeq j_{\rm s}(\lambda_T)^2+2 e_{\rm deg} c_1(\rho_\infty)
\ln(\lambda/\lambda_T) + 2 c_1(\rho_\infty) c_2(\rho_\infty)
(1/\lambda_T-1/\lambda).$$ Here, $\lambda_T$ denotes a crossover
value beyond which the asymptotic behavior becomes valid.  Therefore,
the current diverges logarithmically as $j_{\rm s}^2 \simeq 2 c_1(\rho_\infty)
e_{\rm deg}\ln \lambda$ for large $\lambda$ and $e_{\rm deg}> 0$.

This behavior of $j_{\rm s}$ has interesting implications for the
steady state ligand concentration: $\lambda(x)$ is characterized by a
singularity which occurs at a position $x^*<0$ that moves towards
$x=0$ as $\lambda(0)$ becomes large.  In the vicinity of $x^*$,
$\lambda(x)$ behaves as
\begin{equation}\label{eq:sing_rob}
\lambda\sim (x-x^*)^{-1}(-\ln{(x-x^*)})^{-1/2}.
\end{equation} 
Note, that the case $e_{\rm deg}=0$ has to be discussed separately.
In this case, the current reaches for large $\lambda$ a finite
maximal value $j_{\rm max}$ and the steady state profile diverges as
\begin{equation}\label{eq:sing_rob_edeg0}
\lambda \simeq c_1(\rho_\infty)/(x-x^*) j_{\rm max}.
\end{equation}

If extracellular diffusion is present, i.e. if $D_0>0$, $D_{\rm
s}(\lambda)$ in Eq. (\ref{eq:stst_eqn_full}) changes its asymptotic
behavior to $D_{\rm s}\simeq c_1(\rho_\infty)/\lambda^2+D_0$. For
large $\lambda>\lambda_D$ with $\lambda_D\simeq
(c_1(\rho_\infty)/D_0)^{1/2}$, Eq. (\ref{eq:stst_eqn_full}) becomes
linear and the steady state solution decays exponentially on a length
scale $\xi_d=\sqrt{D_0/e_{\rm deg}}$. The nonlinear behavior described
by Eqs. (\ref{eq:sing_rob}) and (\ref{eq:sing_rob_edeg0}) is thus
valid for $\lambda_T<\lambda<\lambda_D$.

\subsection{Robustness of morphogen gradients}\label{sec:robu}

To study the robustness of morphogen gradients with respect to changes
of the morphogen secretion rate, we consider the response of the
steady state gradient to changes of $j_0$. We define the following
dimensionless measure of robustness:
\begin{equation}\label{eq:defrob}
{\mathcal{R}}(j_0,\lambda)=a (j_0 \partial_{j_0}x(\lambda))^{-1},
\end{equation}
where $x(\lambda)$ is the position at which the steady state ligand
profile attains the concentration $\lambda$.  Here, a robustness of
${\mathcal{R}}(\lambda)=1$ implies that under a 100\% increase of
$j_0$ the position at which the ligand profile attains the fixed value
$\lambda$ is displaced by about one cell diameter $a$, see
Fig.~\ref{fig:robu_illu}. Thus for ${\mathcal{R}}(\lambda)\geq1$, the
shift of the position $x$ where the ligand concentration has the value
$\lambda$ cannot be detected by the cells in the target tissue even
under significant changes of $j_0$.

In absence of extracellular diffusion, the singular behavior
(\ref{eq:sing_rob}) of the steady state profile near $x=x^*$ has
remarkable consequences for the robustness of gradient formation.
Using the robustness ${\mathcal{R}}$ defined in Eq. (\ref{eq:defrob}), we
see from Eq. (\ref{eq:stst}) that ${\mathcal{R}}$ is independent of
$\lambda$.  For steady state equations of the general form
(\ref{eq:stst_eqn_full}), the robustness can be calculated as
\begin{equation}
{\mathcal{R}} = a (j_0 \partial_{j_0}x)^{-1} =  a
\partial_{\lambda_0}j_0/D_{\rm s}(\lambda_0) = a k_{\rm s}(\lambda_0) \lambda_0/j_0
\end{equation}
where $\lambda_0=\lambda(x=0)$ and Eqs. (\ref{eq:stst}) and
(\ref{eq:stst_current}) have been used. The robustness is thus
completely determined by the ratio of the effective degradation rate
and the ligand current at $x=0$ and does not depend on $\lambda$,
i.e. it is the same at all positions $x$ of the concentration gradient
$\lambda(x)$. High degradation rates and small currents lead to a
robust gradient.  Using the asymptotic behavior of the steady state
profile for $D_0=0$, we find that the robustness increases rapidly for
large currents $j_0$ as ${\mathcal{R}}\sim j_0^{-1} e^{j^2_0/j^2_c}$
with $j^2_c = 1/2 c_1(\rho_\infty) e_{\rm deg}$.  For small $j_0$,
${\mathcal{R}}\simeq a/\xi$ becomes constant.  In
Fig.~\ref{fig:robu_illu}~A,B, we illustrate the behavior of the
robustness of steady state gradients for a small and a large value of
$j_0$.

The situation is different if free diffusion in the extracellular
space is present. As discussed in the previous section, the
singularity in the steady state solution disappears for $D_0>0$.  As a
result of this, the robustness approaches a finite value
${\mathcal{R}}_{\rm max} = a/\xi_d$ as $j_0\to\infty$.  In
Fig.~\ref{fig:robu_illu}~D, ${\mathcal{R}}(j_0)$ is shown for
different values of $\xi_d/a$ and Fig.~\ref{fig:robu_illu}~C shows an
example for the effect of the presence of extracellular diffusion on
the robustness of the gradient.

In summary, we find that morphogen gradients can be extremely robust
to changes in the morphogen secretion rate of the source cells if
transport is dominated by transcytosis. The presence of extracellular
diffusion reduces this robustness that is completely lost when
extracellular diffusion is the dominant transport mechanism.

\section{Morphogen transport in two dimensions} \label{sec:2d_desc}

\subsection{Ligand kinetics on the cellular scale}\label{sec:2d_discrete}
The theoretical framework introduced in the previous sections can be
extended to higher dimensions.  Considering a two dimensional
geometry, we represent the cells in the tissue on the sites of a
discrete lattice. The tissue geometry of the wing disk can be captured
by an irregular tiling of the plane, see
Fig.~\ref{fig:lattices_2d}~A,B. For simplicity, however, we use a
triangular lattice with $N$ hexagonal cells in our description, see
Fig.~\ref{fig:lattices_2d}~D. Each hexagonal cell $n=1,\dots,N$ has
$J$ edges $(n,j)$ with $j=1,\dots,J$ ($J=6$ for a triangular lattice)
along which it faces a uniquely defined neighboring cell $n'$ at its
edge $(n',j')$, see Fig.~\ref{fig:lattices_2d}~C. The space between
these two facing edges of the neighboring cells is denoted by the
symbol $\langle n,j \rangle$ with the property $\langle n,j
\rangle=\langle n',j' \rangle$.

To keep our notation simple, we first discuss the case of constant
surface receptor concentration, see section
\ref{sec:const_r}. Assuming that there is no directional bias of
ligand transport, the externalization of the ligand-receptor complexes
can occur on all surfaces of the cell with equal probability. The
equations for the ligand kinetics corresponding to
Eq. (\ref{eq:disc_const_r}) on a lattice are
\begin{eqnarray}
\frac{d}{dt} L_{\langle n,j \rangle} & = & k_{\rm off} ( S_{n,j}+S_{n',j'})
- \frac{J k_{\rm on}}{2}
\left(\frac{2R}{J}- S_{n,j}- S_{n',j'}\right) L_{\langle n,j \rangle} - e_{\rm deg} L_{\langle n,j \rangle}\cr
\frac{d}{dt} S_{n,j} & = & \frac{b_{\rm ext}}{J} S_n^{\rm (i)} - (b_{\rm int}+k_{\rm
  off}) S_{n,j} + \frac{J k_{\rm on}}{2} \left(\frac{R}{J}-S_{n,j}\right)
L_{\langle n,j\rangle}\cr 
\frac{d}{dt} S_n^{\rm (i)} & = & - b_{\rm ext} S_n^{\rm (i)} + b_{\rm int}
\sum_{j=1}^J S_{n,j} - b_{\rm deg} S_n^{\rm (i)},
\label{eq:disc_const_r_2d}
\end{eqnarray}
where $S_n^{\rm (i)}$ is the internal bound ligand concentration in
cell $n$, $S_{n,j}$ the surface bound ligand concentration on edge $j$
of cell $n$, and $L_{\langle n,j \rangle}$ the free ligand
concentration in the extracellular space $\langle n,j \rangle$ which
is located between the two adjacent cells $n$ and $n'$ and the edges
$(n,j)$ and $(n',j')$, see Fig.~\ref{fig:lattices_2d}~C,D.  In
Eq. (\ref{eq:disc_const_r_2d}), we have for simplicity neglected
extracellular diffusion which would couple the concentration
$L_{\langle n,j \rangle}$ to the one on neighboring sites.  At the
boundaries of the lattice, the equation for $\frac{d}{dt} L_{\langle
n,j \rangle}$ in (\ref{eq:disc_const_r_2d}) takes into account ligand
influx analogously to the one dimensional situation, see
Eqs. (\ref{eq:source_bound_disc}) and (\ref{eq:wall_bound_disc}).

\subsection{Transport equations on larger scales}\label{sec:2d_cont}

The effective behavior of ligand transport as described by
Eq. (\ref{eq:disc_const_r_2d}) exhibits anisotropy on large scales due
to the anisotropic lattice structure. We consider for simplicity an
isotropic continuum limit. This simplification is motivated by the
irregular arrangement of cells in a tissue which does not exhibit
lattice anisotropies.

In the isotropic case, the continuum limit describing transport by
transcytosis in two dimensions is of the general form
\begin{equation}\label{eq:dldt_rconst_2d}
\partial_t \lambda = \nabla\cdot (D^{\rm 2d}(\lambda) \nabla \lambda)  - k^{\rm 2d}(\lambda) \lambda
\end{equation}
The effective coefficients $D^{\rm 2d}(\lambda)$ and $k^{\rm
2d}(\lambda)$ in Eq. (\ref{eq:dldt_rconst_2d}) are in general
different from those in the one dimensional case
(\ref{eq:dk_constr_expl}).  In order to determine values for the
coefficients $D^{\rm 2d}(\lambda)$ and $k^{\rm 2d}(\lambda)$ we first
consider concentration profiles which vary only along one symmetry
axis of the lattice given by the $x$-axis in
Fig.~\ref{fig:lattices_2d}~D. In this situation the problem can be
represented on a one dimensional lattice similar to the one
dimensional chain discussed above. We thus determined $D^{\rm
2d}(\lambda)$ and $k^{\rm 2d}(\lambda)$ along this lattice axis using
our one dimensional approach. Lattice symmetry implies that these
coefficients apply to three different lattice axes. In our isotropic
simplification, we assume that they apply to all directions. From this
argument, we find the same effective degradation rate $k^{\rm
2d}(\lambda)=k(\lambda)$ as in one dimension, see
Eq. (\ref{eq:dk_constr_expl}).  The effective diffusion coefficient
changes by a factor of $2/3$: $D^{\rm 2d}(\lambda) = (2/3)
D(\lambda)$.

By the same considerations, the more general description with receptor
kinetics introduced in one dimension in
Eqs. (\ref{eq:lambda_ptc_eq_full}) and (\ref{eq:rho_ptc_eq_full})
generalizes in two dimensions to
\begin{eqnarray}
\partial_t \lambda & = & \nabla\cdot \left(D^{\rm 2d}_\lambda(\lambda, \rho)\nabla\lambda +D^{\rm 2d}_\rho(\lambda, \rho)\nabla\rho\right)
- k^{\rm 2d}_\lambda(\lambda, \rho) \lambda, \cr
\partial_t \rho & = & \nu^{\rm 2d}_{\rm syn}(\lambda,\rho) - k^{\rm 2d}_\rho(\lambda,
\rho) \rho.\nonumber
%\label{eq:dldt_recdyn_2d}
\end{eqnarray}
Here the coefficients $D^{\rm 2d}_\lambda(\lambda, \rho)$, $D^{\rm
2d}_\rho(\lambda, \rho)$, $k^{\rm 2d}_\lambda(\lambda, \rho)$, and
$k^{\rm 2d}_\rho(\lambda, \rho)$ are in general modified due to the
lattice geometry. For a triangular lattice $D^{\rm
2d}_\lambda(\lambda, \rho)=(2/3) D_\lambda(\lambda, \rho)$ and
$D_\rho^{\rm 2d}(\lambda, \rho)=(2/3) D_\rho(\lambda, \rho)$.  The
degradation rates $k^{\rm 2d}_\lambda$ and $k^{\rm 2d}_\rho$ are
identical to those in the one dimensional case. Finally, the rate of
receptor synthesis $\nu^{\rm 2d}_{\rm syn}=\nu_{\rm syn}/a$ is the
same as in one dimension, see Eq. (\ref{eq:nu_1d}), but measured in
different units.  At the boundary line at $x=0$, the ligand source is
described by a current $j_0=\nu/a$ across this boundary line.

We have compared solutions of the effective continuum equation in two
dimensions (\ref{eq:dldt_rconst_2d}) to those of the discrete
description on a triangular lattice (\ref{eq:disc_const_r_2d}). Here,
the ligand source located at $x<0$ extends along the $y$-direction.
This setup is translation invariant along this direction if no
inhomogeneities are present in the tissue. In this situation, the
solutions of the discrete and continuum descriptions are in good
agreement. In order to test the validity of the continuum description
for a case that is not translation invariant along the $y$-direction,
we compared the solutions of Eq. (\ref{eq:dldt_rconst_2d}) for a
geometry in which a rectangular region which the ligand cannot enter
is present in the tissue to solutions of
Eq. (\ref{eq:disc_const_r_2d}) where such a rectangular region is
approximated, see Fig.~\ref{fig:2dcont}.  In the discrete description,
we imposed this constraint by setting $b_{\rm int}=0$ in this
region. In the continuum description (\ref{eq:dldt_rconst_2d}), this
corresponds to $D^{\rm 2d}=0$ within this region which we realized by
imposing a zero flux boundary condition ($j=0$) at its border.  As in
one dimension, the continuum description is appropriate as long as the
degradation rates and the hopping rate $D_0/a^2$ which describes
extracellular diffusion remain small.

\section{Discussion}

In this article, we have first presented a description of morphogen
transport in which cells are discrete entities. This description is
based on key processes like the diffusion of morphogens in the
extracellular space, binding and un-binding of the morphogens to and
from receptor molecules that are located on the cell surfaces,
internalization of these receptor-ligand complexes into the cell and
their subsequent recycling, as well as degradation of external and
internalized ligands, see Fig.~\ref{fig:ptc_scheme}. Moreover, the
production and intracellular trafficking of free receptor molecules by
the cells is included. We have derived effective nonlinear transport
equations (\ref{eq:lambda_ptc_eq_full}) and (\ref{eq:rho_ptc_eq_full})
for the total morphogen concentration and the total receptor
concentration which describe transport by transcytosis on larger
length scales. The effective diffusion coefficient and the effective
degradation rate in these equations are concentration-dependent. If
transcytosis has a directional bias, an additional drift term appears
in the transport equations.

Other mechanisms of ligand transport can be effectively described by
equations (\ref{eq:lambda_ptc_eq_full}) and
(\ref{eq:rho_ptc_eq_full}).  The effective transport coefficients can
be derived from a detailed description of any particular transport
mechanism. As an example for this, we have discussed a model of
morphogen transport where transport occurs via diffusion of ligands
bound to carrier molecules in the cell membrane \cite{bele04,han04},
see appendix \ref{sec:glyp}.

Our theoretical description of morphogen transport captures the key
processes that are relevant for ligand transport and ligand-receptor
interactions in multicellular epithelia such as the wing disk. We used
a simplified description of these processes and neglected several
aspects that could play a role. For example, we did not account for
cell divisions and tissue growth in our description \cite{shra05}.
More importantly, the presence of different receptor types which is
quite common for signaling molecules of the TGF-$\beta$ superfamily
like Dpp was neglected in our description.  These receptors typically
form dimers or other complexes and, in general, the affinity of the
ligand is different for the different receptor types and
complexes. The trafficking of ligand-receptor pairs inside the cell is
a very complex process that is only crudely captured in our
description by a few parameters.

Our coarse-graining procedure starts from a discrete cellular
representation and allows us to obtain effective transport equations
in a continuum limit. This provides a theoretical framework for a
quantitative analysis of the spreading and trafficking of signaling
molecules in and between cells. Using this approach one can relate
experimental data obtained at different scales ranging from the cell
to the tissue level.  For example, the situation shown in
Fig.~\ref{fig:2dcont} mimics recent experiments done in the Drosophila
wing disk in which endocytosis is blocked in patches of cells in the
tissue (see section \ref{sec:wing_disk}) \cite{entc00}. The
calculation results shown in Fig.~\ref{fig:2dcont} are consistent with
the experimental data obtained in these shibire clone experiments
\cite{entc00}. They show a ligand depletion of decreasing relative
magnitude (``contrast'') behind the clone region as it is observed
experimentally, see Fig.~\ref{fig:2dcont}~H. Note, that the clone
region itself is devoid of ligands which is also seen experimentally
and is evidence for a transport mechanism via transcytosis
\cite{krus04}.

Transport processes of signaling molecules in tissues show many
features in common which are captured by our description. Our study
highlights some general properties of these systems such as the
robustness of gradients which are largely independent of parameter
values and molecular details. We expect that such general features can
play an important role in very different biological signaling systems.

\appendix
\section{Derivation of  the effective transport
  equations in one dimension}
  \label{sec:deri}

To perform a continuum limit of Eq. (\ref{eq:ptc_full_disc}), we introduce
the densities $l(t,x)$, $r^{\rm (l/r)}(t,x)$, $r_i(t,x)$, $s^{\rm
(l/r)}(t,x)$, and $s_i(t,x)$, such that $x=na,\ L_n(t)/a = l(t,x) ,\
R_n^{\rm (l)}(t)/a = r^{\rm (l)}(t,x),\ R_n^{\rm (r)}(t)/a = r^{\rm
(r)}(t,x),\ R_n^{\rm (i)}/a = r_i(t,x),\ S_n^{\rm (l)}(t)/a = s^{\rm
(l)}(t,x),\ S_n^{\rm (r)}(t)/a = s^{\rm (r)}(t,x),\ \textnormal{and }\
S_n^{\rm (i)}/a = s_i(t,x)$. Kinetic equations for these are obtained
by replacing the discrete densities $R_n^{\rm (i)}$, $R_n^{\rm
(l/r)}$, $S_n^{\rm (i)}$, $S_n^{\rm (l/r)}$, and $L_n$ in
Eq. (\ref{eq:ptc_full_disc}) with the continuum densities $r_i$, $r^{\rm
(l/r)}$, $s_i$, $s^{\rm (l/r)}$, and $l$. The spatial separation of
the quantities defined on the lattice as indicated in
Fig.~\ref{fig:ptc_scheme} is taken into account by including terms
up to second order in a power series expansion in $x$. We also change
variables to $r_\pm(t,x) = r^{\rm (l)}(t,x)\pm r^{ (r)}(t,x)$ and
$s_\pm(t,x)=s^{\rm (l)}(t,x)\pm s^{\rm (r)}(t,x)$ so that $r_+$ and
$s_+$ measure the total free and ligand bound surface receptor
concentrations per cell and $r_-$ and $s_-$ the polarization of these
concentrations on the cell surface respectively. This yields the
continuum equations
\begin{eqnarray}\label{eq:contlim1}
\partial_t l & = & k_{\rm off} s_+ - (a k_{\rm on} r_+ +e_{\rm deg})l - \frac{ab^2 k_{\rm on}}{8}\, l \partial_x^2r_+ + \frac{b^2 k_{\rm off}}{8}\,
\partial_x^2s_+ + D_0\, \partial_x^2 l  \cr
& &-\frac{ab k_{\rm on}}{2}\, l \partial_xr_- 
+ \frac{b k_{\rm off}}{2}\, \partial_x s_- \cr
\partial_t s_+ & = & a k_{\rm on}l r_+ + b_{\rm ext} s_i - (b_{\rm int}+k_{\rm off}) s_+ + \frac{ab^2 k_{\rm on}}{8}\, r_+ \partial_x^2l
+\frac{(a-b)^2 b_{\rm ext}}{8}\, \partial_x^2 s_i - \frac{ab k_{\rm on}}{2}\, r_- \partial_x l \cr
\partial_t s_i & = & -(b_{\rm ext}+b_{\rm deg})s_i+b_{\rm int} s_+ + \frac{(a-b)^2 b_{\rm int}}{8}\, \partial_x^2 s_+ - 
\frac{(a-b) b_{\rm int}}{2}\, \partial_x s_-\cr
\partial_t s_- & = & a k_{\rm on} l r_- - (b_{\rm int}+k_{\rm off}) s_- + \frac{ab^2 k_{\rm on}}{8}\, r_- \partial_x^2 l -
\frac{ab k_{\rm on}}{2}\, r_+ \partial_x l + \frac{(a-b) b_{\rm ext}}{2}\, \partial_x s_i\cr
\partial_t  r_+ & = & \frac{f_{\rm syn}}{a} + f_{\rm ext} r_i - f_{\rm int} r_+ - ak_{\rm on}l r_+ + k_{\rm off} s_+ 
-\frac{(ab^2) k_{\rm on}}{8} r_+ \partial_x^2l\cr
& &+\frac{(a-b)^2 f_{\rm ext}}{8}\, \partial_x^2 r_i  
+ \frac{ab k_{\rm on}}{2}\, r_- \partial_x l\cr
\partial_t  r_i & = & -(f_{\rm ext}+f_{\rm deg}) r_i + f_{\rm int} r_+ + \frac{(a-b)^2 f_{\rm int}}{8}\, \partial_x^2 r_+
-\frac{(a-b) f_{\rm int}}{2}\, \partial_x r_- \cr
\partial_t  r_- & = & - f_{\rm int} r_- - a k_{\rm on} l r_- + k_{\rm off} s_- - \frac{ab^2 k_{\rm on}}{8}\, r_- \partial_x^2 l
+\frac{ab k_{\rm on}}{2}\, r_+ \partial_x l+\frac{(a-b) f_{\rm ext}}{2}\, \partial_x r_i \label{eq:fullcont}.
\end{eqnarray}
Here, it is reasonable to neglect derivatives of higher order with
respect to $x$ because the most important contribution to ligand
transport on large length scales comes from the second derivative
terms. This is due to the fact that first derivative terms must not
appear in an effective transport equation because of the mirror
symmetry of the original description.

In the absence of degradation and production, there are two conserved
quantities in the system, namely the total ligand number and the total
receptor number. Indeed, the kinetic equation for the total ligand
density $\lambda=l+s_i+s_+$ that follows from Eq. (\ref{eq:fullcont}) can
be written as a continuity equation with sink term
\begin{equation} \label{eq:dlambdadt}
  \partial_t \lambda  =  -\partial_x j - b_{\rm deg} s_i - e_{\rm deg} l. 
\end{equation}
Here, the total ligand current is 
\begin{eqnarray}
j  & = & \frac{ab k_{\rm on}}{2} l r_- + \frac{(a-b) b_{\rm int}}{2} s_- - \frac{b k_{\rm off}}{2} s_-
- \frac{ab^2 k_{\rm on}}{8} \left(r_+ \partial_x l - l \partial_x r_+\right) \cr
& & - \frac{(a-b)^2 b_{\rm ext}}{8} \partial_x s_i 
- \frac{(a-b)^2 b_{\rm int} + b^2 k_{\rm off}}{8} \partial_x s_+ - D_0
\partial_x l. \label{eq:tot_lig_j}
\end{eqnarray}
Note, that the terms involving $r_-$ and $s_-$ appear directly whereas
all other terms are proportional to derivatives of $r_+$, $s_+$,
$s_i$, or $l$.  The kinetics of the total receptor density $\rho =
r_i + r_+ + s_i + s_+$ is given by another continuity equation with
source and sink terms:
\begin{equation}
  \partial_t \rho  =  - \partial_x j_\rho + \frac{f_{\rm syn}}{a} - f_{\rm deg} r_i
  - b_{\rm deg} s_i  \label{eq:drhodt}
\end{equation}
with the total receptor current
\begin{equation}
j_\rho  =  \frac{(a-b)}{2} ( f_{\rm int} r_- +  b_{\rm int} s_-) -
 \frac{(a-b)^2}{8} \left(f_{\rm ext} \partial_x r_i + 
 f_{\rm int} \partial_x r_+ 
+ b_{\rm ext} \partial_x s_i + b_{\rm int} \partial_x s_+\right).\label{eq:tot_rec_j}
\end{equation}
The individual terms of the currents $j$ and $j_\rho$ are difficult to
interpret.  However, it will become clear below that the terms in
$j_\rho$ do not give rise to transport over large distances whereas
this is the case for the terms in $j$.

The equations (\ref{eq:dlambdadt}) and (\ref{eq:drhodt}) have the
unpleasant property that they relate the time development of $\lambda$
and $\rho$ to that of all the individual quantities
$l,s_i,s_\pm,r_i,r_\pm$ whose time development is given by the set of
coupled partial differential equations (\ref{eq:contlim1}). It would
be better if the kinetics of $\lambda$ and $\rho$ could be described
by equations which only involve these two quantities. This can be
achieved by exploiting a separation of time scales.

As discussed in the main text, the relaxation time scale $\tau_a$ for
the kinetics in one cell is much smaller than the time scale
$\tau_{\xi_D}$ for ligand transport on a large length scale $\xi_D$ on
which the ligand profile develops.  As we are interested in the
behavior on large length scales in the continuum description, we
exploit $\tau_a \ll \tau_{\xi_D}$ by making an adiabatic approximation
in which the system equilibrates infinitely fast locally.  This is
done by setting all time derivatives in Eq. (\ref{eq:fullcont}) to
zero and neglecting the second derivative terms.  The resulting
equations provide 5 relations between the 7 variables $l, r_i, r_\pm,
s_i, s_\pm$. Here, we also assume that the production and degradation
rates are small compared to the other rates. Formally, this
corresponds to setting $b_{\rm deg}=e_{\rm deg}=f_{\rm deg}=f_{\rm
syn}=0$.  This procedure yields the relations (\ref{eq:ad_ap_f_1d}).

Note, that we have kept the first derivative terms for $r_-$ and $s_-$
in Eq. (\ref{eq:ad_ap_f_1d}). This is done to retain all second
derivative terms when inserting Eq. (\ref{eq:full_le}) into
Eqs. (\ref{eq:dlambdadt}), (\ref{eq:tot_lig_j}), (\ref{eq:drhodt}),
and (\ref{eq:tot_rec_j}).  Using Eq. (\ref{eq:full_le}), one can
express $l, r_i, r_-, s_i, s_-$ in terms of $r_+$ and $s_+$ and
spatial derivatives thereof. Finally, $r_+$ and $s_+$ can be expressed
in terms of $\rho=r_i+r_++s_i+s_+$ and $\lambda=l+s_i+s_+$.
Mathematically, there exist two solutions for $s_+(\lambda,\rho)$ and
$r_+(\lambda,\rho)$ but only one of them satisfies the physical
requirement that $s_+(\lambda=0,\rho)=0$ and $r_+(\lambda,\rho=0)=0$.
Thus, we can uniquely express $l, r_i, r_\pm, s_i$, and $s_\pm$ in
terms of $\lambda$ and $\rho$ in the adiabatic approximation.

Using these expressions, it is straightforward to cast
Eqs. (\ref{eq:dlambdadt}), (\ref{eq:tot_lig_j}), (\ref{eq:drhodt}),
and (\ref{eq:tot_rec_j}) into the two coupled partial differential
equations (\ref{eq:lambda_ptc_eq_full}) and
(\ref{eq:rho_ptc_eq_full}).  The explicit expressions for the
transport and degradation coefficients in this one dimensional
description of morphogen transport are
\begin{eqnarray}
D_\lambda(\lambda, \rho) & = & - {a^3{b_{\rm ext}}{b_{\rm int}}( {b_{\rm ext}} + {b_{\rm int}} ) {{f_{\rm ext}}}^2{f_{\rm int}}{k_{\rm off}}
      {k_{\rm on}}\rho} [-2{b_{\rm int}} ( {b_{\rm ext}}( {f_{\rm
      ext}} + {f_{\rm int}} ) {k_{\rm off}} \cr
& &+ a( {b_{\rm ext}} + {b_{\rm int}} ) {f_{\rm ext}}{k_{\rm on}}\lambda )^2 
+   2A(\lambda,\rho)( -( {b_{\rm int}}{f_{\rm ext}}( 2{f_{\rm int}}( {b_{\rm int}} + {k_{\rm off}} )  + 
              a{b_{\rm int}}{k_{\rm on}}\lambda )  )  \cr
& &+{b_{\rm ext}}( -2{f_{\rm ext}}{f_{\rm int}}{k_{\rm off}} + 
            {b_{\rm int}}( {f_{\rm int}}{k_{\rm off}} 
+    {f_{\rm ext}}( -2{f_{\rm int}} + {k_{\rm off}} - a{k_{\rm
      on}}\lambda )  )  )  )  \cr
& &+ 2a{b_{\rm int}}( {b_{\rm ext}} + {b_{\rm int}} ) {f_{\rm ext}}{k_{\rm on}}
       ( A(\lambda,\rho) - 2{b_{\rm ext}}( {f_{\rm ext}} + {f_{\rm
      int}} ) {k_{\rm off} } \cr
& & +   2a( {b_{\rm ext}} + {b_{\rm int}} ) {f_{\rm ext}}{k_{\rm on}}\lambda ) \rho - 
      2a^2{b_{\rm int}}{( {b_{\rm ext}  } + {b_{\rm int}} )
      }^2{{f_{\rm ext}}}^2{{k_{\rm on} }}^2{\rho}^2]^{-1} \cr
& & + 2 D_0 { a {b_{\rm ext}} ({b_{\rm ext}}+{b_{\rm int}}) {f_{\rm ext}}
      ({f_{\rm ext}}+{f_{\rm int}}) {k_{\rm off}} {k_{\rm on}} \rho } [A(\lambda,\rho)
      (A(\lambda,\rho)\cr
& & -{b_{\rm ext}} ({f_{\rm int}}   {k_{\rm off}}+{f_{\rm ext}} ({k_{\rm off}}+a {k_{\rm on}} (\lambda -\rho
      )))+a {b_{\rm int}} {f_{\rm ext}} {k_{\rm on}} (\rho -\lambda ))]^{-1} \cr
%%%%%%%%%%%%%%%%%%%%%%%%%%%%%%%%%%%%%%%%%%%%%%%%%%%%%%%%%%%%%%%%%%%%%%%%%%%%%%%%%%%%%%%%%%%%%%%%%%
D_\rho(\lambda, \rho) & = & {a^4{b_{\rm ext}}{b_{\rm int}}( {b_{\rm ext}} + {b_{\rm int}} ) {{f_{\rm ext}}}^2{f_{\rm int}}{k_{\rm off}}
    {{k_{\rm on}}}^2\lambda\rho} [2( ( -( {f_{\rm int}}( {b_{\rm int}} + {k_{\rm off}} )  )  - 
         a{b_{\rm int}}{k_{\rm on}}\lambda ) \cr
& & ( {b_{\rm ext}}( {f_{\rm ext}} + {f_{\rm int}} ) {k_{\rm off}} + 
         a( {b_{\rm ext}} + {b_{\rm int}} ) {f_{\rm ext}}{k_{\rm
    on}}\lambda )   
( A(\lambda,\rho) \cr
& & + {b_{\rm ext}}( {f_{\rm ext}} + {f_{\rm int}} ) {k_{\rm off}} + 
         a( {b_{\rm ext}} + {b_{\rm int}} ) {f_{\rm ext}}{k_{\rm on}}\lambda )\cr
& &  +    a( {b_{\rm ext}} + {b_{\rm int}} ) {f_{\rm ext}}{k_{\rm on}} 
       ( -( A(\lambda,\rho){f_{\rm int}}( {b_{\rm int}} + {k_{\rm
    off}} )  )  + aA(\lambda,\rho){b_{\rm int}}{k_{\rm on}}\lambda \cr
& &+  2( {f_{\rm int}}( {b_{\rm int}} + {k_{\rm off}} )  + a{b_{\rm int}}{k_{\rm on}}\lambda ) 
   ( -( {b_{\rm ext}}( {f_{\rm ext}} + {f_{\rm int}} ) {k_{\rm off}} )
    \cr
& & +     a( {b_{\rm ext}} + {b_{\rm int}} ) {f_{\rm ext}}{k_{\rm
    on}}\lambda )  ) \rho 
 -   a^2{( {b_{\rm ext}} + {b_{\rm int}} ) }^2{{f_{\rm ext}}}^2{{k_{\rm on}}}^2\cr
& & \times ( {f_{\rm int}}( {b_{\rm int}} + {k_{\rm off}} )  +
    a{b_{\rm int}}{k_{\rm on}}\lambda ) {\rho}^2 ) ]^{-1} \cr
& & + 2 D_0  {b_{\rm ext}} ({f_{\rm ext}}+{f_{\rm int}}) {k_{\rm
    off}}  ( ((({f_{\rm ext}}+{f_{\rm int}}) {k_{\rm off}}+a
    {f_{\rm ext}} {k_{\rm on}} \lambda )^2\cr
& & +a {f_{\rm ext}}
   {k_{\rm on}} (({f_{\rm ext}}+{f_{\rm int}}) {k_{\rm off}}-a {f_{\rm
    ext}} {k_{\rm on}} \lambda ) \rho  ) {b_{\rm ext}}^2\cr
& & -A(\lambda,\rho) (({f_{\rm ext}}+{f_{\rm int}}) {k_{\rm off}}+a
   {f_{\rm ext}} {k_{\rm on}} \lambda ) {b_{\rm ext}}+a {b_{\rm int}}
    {f_{\rm ext}} {k_{\rm on}} (2 {f_{\rm ext}} \lambda  ({k_{\rm
    off}}+a {k_{\rm on}} \lambda )\cr
& & +{f_{\rm ext}}
   ({k_{\rm off}}-2 a {k_{\rm on}} \lambda ) \rho +{f_{\rm int}}
    {k_{\rm off}} (2 \lambda +\rho )) {b_{\rm ext}}+a {b_{\rm int}}
    {f_{\rm ext}} {k_{\rm on}} \lambda  \cr 
& & \times (a {b_{\rm int}}
   {f_{\rm ext}} {k_{\rm on}} (\lambda -\rho
    )-A(\lambda,\rho)) ) [A(\lambda,\rho) (A(\lambda,\rho)-{b_{\rm
    ext}} ({f_{\rm int}} {k_{\rm off}}\cr
& & +{f_{\rm ext}} ({k_{\rm off}}+a {k_{\rm on}} (\lambda -\rho )))+a
   {b_{\rm int}} {f_{\rm ext}} {k_{\rm on}} (\rho -\lambda ))^2]^{-1} \cr
%%%%%%%%%%%%%%%%%%%%%%%%%%%%%%%%%%%%%%%%%%%%%%%%%%%%%%%%%%%%%%%%%%%%%%%%%%%%%%%%%%%%%%%%%%%%%%
k_\lambda(\lambda, \rho) & = & [{b_{\rm deg}}{b_{\rm int}}( a{b_{\rm
      int}}{f_{\rm ext}}{k_{\rm on}}( \lambda + \rho )  - 
      {A(\lambda, \rho)} + {b_{\rm ext}}
       ( {f_{\rm int}}{k_{\rm off}} \cr
& &+ {f_{\rm ext}}( {k_{\rm off}} + a{k_{\rm on}}( \lambda + \rho )  )  ) 
      ) ] [{2a{( {b_{\rm ext}} + {b_{\rm int}} ) }^2{f_{\rm ext}}{k_{\rm on}}\lambda}]^{-1}\cr
%%%%%%%%%%%%%%%%%%%%%%%%%%%%%%%%%%%%%%%%%%%%%%%%%%%%%%%%%%%%%%%%%%%%
k_\rho(\lambda, \rho) & = & [\left( -\left( \left( {b_{\rm ext}} + {b_{\rm int}} \right) {f_{\rm deg}}{f_{\rm int}} \right)  + 
       {b_{\rm deg}}{b_{\rm int}}\left( {f_{\rm ext}} + {f_{\rm int}} \right)  \right) 
     ( -A(\lambda,\rho) \cr
& &+ {b_{\rm ext}}\left( {f_{\rm ext}} + {f_{\rm int}} \right)
       k_{\rm off} 
       +  a\left( {b_{\rm ext}} + {b_{\rm int}} \right) {f_{\rm
       ext}}{k_{\rm on}}\lambda )  \cr
& &+  a\left( {b_{\rm ext}} + {b_{\rm int}} \right) {f_{\rm ext}}
     \left( \left( {b_{\rm ext}} + {b_{\rm int}} \right) {f_{\rm deg}}{f_{\rm int}} + 
       {b_{\rm deg}}{b_{\rm int}}\left( {f_{\rm ext}} + {f_{\rm int}} \right)  \right)
       {k_{\rm on}}\rho]\cr
& & \times [2a
    {\left( {b_{\rm ext}} + {b_{\rm int}} \right) }^2{f_{\rm
       ext}}\left( {f_{\rm ext}} + {f_{\rm int}} \right) {k_{\rm
       on}}\rho]^{-1}, {\rm with}\cr
%%%%%%%%%%%%%%%%%%%%%%%%%%%%%%%%%%%%%%%%%%%%%%%%%%%%%%%%%%%%%%%%%%%%%%%%%%%%%%%%%%
& &  \cr
A(\lambda, \rho) & = & [{( {b_{\rm ext}}( {f_{\rm int}}{k_{\rm off}} + 
                 {f_{\rm ext}}( {k_{\rm off}} + a{k_{\rm on}}( \lambda - \rho )  )  )  + 
              a{b_{\rm int}}{f_{\rm ext}}{k_{\rm on}}( \lambda - \rho )  ) }^2 \cr
& & +   4a{b_{\rm ext}}( {b_{\rm ext}} + {b_{\rm int}} ) {f_{\rm ext}}(
                 {f_{\rm ext}} + {f_{\rm int}} ) 
           {k_{\rm off}}{k_{\rm on}}\rho]^{1/2} .
\label{eq:dk_expl_1d}
\end{eqnarray}

\section{Directional bias of intracellular trafficking}\label{sec:ptc_bias}

We study the effects of a bias in the description of transcytosis.  We
introduce a dimensionless parameter $-1/2\leq \beta\leq 1/2$ which
measures this bias in Eq. (\ref{eq:ptc_full_disc}). The kinetic equations
of the discrete description with bias are:
\begin{eqnarray}\label{eq:ptc_bias_disc}
\frac{d}{dt} S_n^{\rm (r)} & = & -k_{\rm off} S_n^{\rm (r)} + k_{\rm on} R_n^{
 \rm (r)}L_{ n} - b_{\rm int} S_{ n}^{\rm (r)} +  b_{\rm ext}(1/2+\beta)\, S_n^{
 \rm (i)} \cr 
\frac{d}{dt} S_n^{\rm (l)} & = & -k_{\rm off} S_n^{\rm (l)} + k_{\rm on} R_n^{
 \rm (l)}L_{ n-1} - b_{\rm int} S_{ n}^{\rm (l)} + b_{\rm ext}(1/2-\beta)\, S_n^{
 \rm (i)} ,
\end{eqnarray}
with the kinetics for the remaining quantities as in
Eq. (\ref{eq:ptc_full_disc}).  For $\beta=-1/2$, all receptor-bound
ligands are externalized on the left surface, for $\beta=1/2$ on the
right surface, and for $\beta=0$, we recover the unbiased description.
This implies that the externalization of the free receptors remains
unbiased.  We proceed as before to derive the continuum equations. The
adiabatic approximation changes to
\begin{equation}
s_-  =  \frac{k_{\rm on} a l r_- - 2\beta b_{\rm ext} s_i - \frac{ab k_{\rm on}}{2}\, r_+ \partial_x l 
+ \frac{(a-b) b_{\rm ext}}{2}\, \partial_x s_i}{b_{\rm int} + k_{\rm off}} 
\end{equation}
with the other relations as in Eq. (\ref{eq:full_le}).  Finally, the
transport equations (\ref{eq:lambda_ptc_eq_full}) and
(\ref{eq:rho_ptc_eq_full}) generalize to
Eq. (\ref{eq:ptc_eq_full_bias}) with the new effective drift velocity
\begin{eqnarray}
V_\beta(\lambda,\rho) & = & 2 a^2 \beta  {b_{\rm ext}}  {b_{\rm int}}  {f_{\rm ext}}  {f_{\rm int}}  {k_{\rm off}}  {k_{\rm on}} \rho
  [ {b_{\rm ext}}  (  {f_{\rm int}}  (  {b_{\rm int}} +  {k_{\rm off}}  )  + a  {b_{\rm int}}  {k_{\rm on}} \lambda  )  
      (  (  {f_{\rm ext}} +  {f_{\rm int}}  )   {k_{\rm off}} \cr
& & + a  {f_{\rm ext}}  {k_{\rm on}} \lambda  )  + 
    a  {b_{\rm ext}}  {f_{\rm ext}}  {k_{\rm on}}  (  {f_{\rm int}}  (  {b_{\rm int}} +  {k_{\rm off}}  )  - 
       a  {b_{\rm int}}  {k_{\rm on}} \lambda  )  \rho \cr
& & + 
     (  {f_{\rm int}}  (  {b_{\rm int}} +  {k_{\rm off}}  )  + a  {b_{\rm int}}  {k_{\rm on}} \lambda  )  
     {A(\lambda,\rho)} + a  {b_{\rm int}}  {f_{\rm ext}}  {k_{\rm on}} 
      (  {f_{\rm int}}  {k_{\rm off}}  ( \lambda + \rho  )  \cr
& & +        {b_{\rm int}}  ( a  {k_{\rm on}} \lambda  ( \lambda - \rho
   )  +  {f_{\rm int}} ( \lambda + \rho  )   )  ) ]^{-1},
\label{eq:vbeta_expl}
\end{eqnarray}
where $A(\lambda,\rho)$ is defined in Eq. (\ref{eq:dk_expl_1d}).  The
other coefficients appearing in Eq. (\ref{eq:ptc_eq_full_bias}) remain as
in the case without directional bias.

\section{Simple limits of the effective transport equations}\label{sec:simp_lim}

In biological systems, it is well possible that some of the processes
included in our description of ligand transport are much faster than
the others. For example, the binding of the ligand to its receptor can
be fast compared to other processes due to the small volume
of the gaps between cells. The confinement of the ligands to this
small volume leads to frequent collisions between ligands and
receptors. Assuming that the reaction is diffusion limited this
can lead to a high reaction rate $k_{\rm on}$.

It is worthwhile to note that Eqs. (\ref{eq:lambda_ptc_eq_full}) and
(\ref{eq:rho_ptc_eq_full}) become simpler if this or another one of
the cellular processes is much faster than the others, i.e. if the
corresponding rates in our description are very large.  For example,
in the limit of very fast binding of ligands to receptors, i.e. for
$k_{\rm on} \to \infty$, we find
\begin{eqnarray}
D_\lambda(\lambda,\rho) & = &
\frac{-a^2{b_{\rm ext}}{b_{\rm int}}{f_{\rm ext}}{f_{\rm int}}{k_{\rm off}}\rho
}{B(\lambda, \rho)} \cr
D_\rho(\lambda,\rho) & = &
\frac{a^2{b_{\rm ext}}{b_{\rm int}}{f_{\rm ext}}{f_{\rm int}}{k_{\rm
      off}}\lambda}{B(\lambda, \rho)} \cr 
k_\lambda(\lambda,\rho) & = & \frac{{b_{\rm deg}}{b_{\rm int}}}{{b_{\rm ext}} + {b_{\rm int}}}\cr
k_\rho(\lambda,\rho) & = & \frac{{b_{\rm deg}}{b_{\rm int}}( {f_{\rm ext}} + {f_{\rm int}} ) \lambda + 
    ( {b_{\rm ext}} + {b_{\rm int}} ) {f_{\rm deg}}{f_{\rm int}}(
  \rho-\lambda   ) }{( {b_{\rm ext}} + 
      {b_{\rm int}} ) ( {f_{\rm ext}} + {f_{\rm int}} ) \rho},
\label{eq:explicit_kon_fast}
\end{eqnarray}
where
\begin{eqnarray}
B(\lambda, \rho) & = & 4( {b_{\rm int}}{f_{\rm ext}}{f_{\rm int}}( {b_{\rm int}} + {k_{\rm off}} )  + 
     {b_{\rm ext}}( {b_{\rm int}}{f_{\rm ext}}{f_{\rm int}} + {f_{\rm
     ext}}{f_{\rm int}}{k_{\rm off}} \cr
& & - {b_{\rm int}}( {f_{\rm ext}} + {f_{\rm int}} ) {k_{\rm off}} )) \lambda 
 -  4( {b_{\rm ext}} + {b_{\rm int}} ) {f_{\rm ext}}{f_{\rm int}}(
  {b_{\rm int}} + {k_{\rm off}} ) \rho. \nonumber
\end{eqnarray}
 There are no free ligands in this limit because $l=0$ via
Eq. (\ref{eq:full_le}). As all ligands are bound to receptors, free
diffusion does not contribute to the current and $D_0$ does not appear
in Eq. (\ref{eq:explicit_kon_fast}). This also imposes the constraint
$\lambda\le\rho$. The example given by
Eq. (\ref{eq:explicit_kon_fast}) is instructive because the effective
transport and degradation coefficients are much simpler than those in
the general case (\ref{eq:dk_expl_1d}).  Many of the properties
discussed above for the general case that is valid for arbitrary
$k_{\rm on}$ can be read directly from the expressions in Eq.
(\ref{eq:explicit_kon_fast}).  For example, if $b_{\rm ext}=0$ ligands
do not move because $D_\lambda=D_\rho=0$ in
Eq. (\ref{eq:explicit_kon_fast}).  Due to the constraint
$\lambda\le\rho$, however, the statements for the asymptotic behavior
for $\lambda \to \infty$ do not apply to this case. Furthermore,
$D_\lambda$ does no longer exhibit a maximum as a function of
$\lambda$.  It either grows or decreases monotonically depending on
the parameter choice.

In principle, one can write down simpler expressions as in
Eq. (\ref{eq:explicit_kon_fast}) for many different limits. If several
transport steps are much faster than the others, only the ratios of
the corresponding parameters enter the simplified description. For
example, if the binding and un-binding of the ligand to the receptor
is faster than all other processes the effective diffusion coefficient
and degradation rate do not depend on $k_{\rm on}$ and $k_{\rm off}$
individually but only on the ratio $k_{\rm on}/k_{\rm off}$.  The
number of parameters can thus be reduced to obtain the minimal
description for a given situation.  A simple but instructive example
is the situation $k_{\rm on}\gg k_{\rm off}\gg b_{\rm ext}\gg b_{\rm
int}$ in Eq. (\ref{eq:dldt_rconst}), for which we obtain the effective
diffusion coefficient
$$ \lim_{b_{\rm ext}\to \infty}\lim_{k_{\rm off}\to
  \infty}\lim_{k_{\rm on}\to \infty} D(\lambda) = b_{\rm int} a^2/4.$$
  This reflects that the only slow process --- in this case the
  internalization of ligands at rate $b_{\rm int}$ --- limits the
  transport efficiency and defines the effective diffusion
  coefficient.  Note, that the limits taken above do not commute
  because $D(\lambda)=0$ for $k_{\rm off}\gg k_{\rm on}$.

\section{Transport by diffusion in the cell  membrane}\label{sec:glyp}

 As another example for an application of our theoretical framework,
consider a transport mechanism in which ligands can move across cells
by passive diffusion in the cell membrane.  The ligand first binds to
a receptor molecule on the cell surface. This complex then diffuses in
the cell membrane. At any time, the ligand can detach from the
receptor it occupies and, after diffusing over a short distance in the
extracellular space, it can attach to a new receptor that can be
located on the surface of the same or a different cell.  Note, that if
we simply replace the receptors with HSPG molecules this transport
mechanism is very similar to one that was recently suggested for Dpp
in the wing disk \cite{bele04}.

We can describe this mechanism on the same lattice structure as used
above for transcytosis, see Fig.~\ref{fig:ptc_scheme}.  For
simplicity, we focus on the essence of the transport phenomenon and do
not include ligand degradation or the production and degradation of
receptors in our description.  Furthermore, the receptor concentration
is assumed to be constant on the cell surface. This description is
similar to that of transcytosis with constant surface receptor
concentration given by Eq. (\ref{eq:disc_const_r}).  However, instead
of the internalization and externalization of receptor-ligand
complexes as in transcytosis, dislocation of the ligand-receptor
complexes across one cell is due to diffusion on the cell surface. In
a discrete one dimensional description, this effect is captured by
hopping at a rate $h\simeq D_{\rm M}/2a^2$ between the left and right
surface of each cell, where $D_{\rm M}$ is the diffusion coefficient
of receptors in the cell membrane. Using the same notation as in
Eq. (\ref{eq:disc_const_r}), a discrete description of this mechanism
reads
\begin{eqnarray}
\frac{d}{dt} L_n & = & k_{\rm off} (S_n^{\rm (r)} + S_{ n+1}^{\rm (l)}) - k_{\rm on}
(R - S_n^{\rm (r)} - S_{ n+1}^{\rm (l)} ) L_n \cr
\frac{d}{dt} S_n^{\rm (r)} & = & -k_{\rm off} S_n^{\rm (r)} + k_{\rm on}
\Bigl(R/2-S_n^{\rm (r)}\Bigr) L_{ n} + h (S_{ n}^{\rm (l)}- S_n^{\rm (r)}) \cr
\frac{d}{dt} S_n^{\rm (l)} & = & -k_{\rm off} S_n^{\rm (l)} + k_{\rm on}
\Bigl(R/2-S_n^{\rm (l)}\Bigr) L_{ n-1} + h (S_{ n}^{\rm (r)}-
S_n^{\rm (l)}).
\nonumber
\end{eqnarray}

Applying the same method as in section \ref{sec:1d_cont}, we obtain
the continuum transport equation $\partial_t \lambda=\partial_x
(D_{\rm SD}(\lambda)\partial_x\lambda)$ where $\lambda$ is the total
ligand concentration and
\begin{eqnarray} 
D_{\rm SD}(\lambda) & = & \frac{a^3 r h {k_{\rm off}} {k_{\rm on}}  
    \left( E(\lambda) - {k_{\rm off}} + a {k_{\rm on}} \left( -\lambda + r  \right)  \right) }{4 E(\lambda) 
    \left( E(\lambda) h - h \left( {k_{\rm off}} + a {k_{\rm on}} \lambda \right)  + 
      a r \left( h + {k_{\rm off}} \right)  {k_{\rm on}}  \right) }
    \textnormal{, with} \cr
E(\lambda) & = & {\left[-4 a^2 r {{k_{\rm on}}}^2 \lambda + {\left(
    {k_{\rm off}} + a {k_{\rm on}} \left( r + \lambda \right)  \right)
    }^2\right]^{1/2}}.
\end{eqnarray}
 $D_{\rm SD}(\lambda)$ has properties that are similar to those of the
effective diffusion coefficient of the transcytosis model. It exhibits
a maximum as a function of $\lambda$ and decays as $D_{\rm SD} \sim
\lambda^{-2}$ for large $\lambda$. Hence, upon inclusion of ligand
degradation in the description, gradient formation by this mechanism
exhibits similar properties as transcytosis.

\section{Transport by extracellular diffusion}\label{sec:theor_dbts}
In order to ultimately identify the mechanism of morphogen gradient
formation present in a given system, it is important to develop
mathematical descriptions of all potentially relevant transport
mechanisms so that these can be compared to the available experimental
data. Extracellular diffusion is widely believed to be the dominant
transport mechanism for some morphogens \cite{gurd94,land02}. For this
reason, we briefly discuss the extracellular diffusion dominated limit
of the transport phenomenon defined by Eq. (\ref{eq:ptc_full_disc}) in
this section.

Assuming that extracellular diffusion gives the dominant contribution
to the ligand current, we neglect the contribution of
transcytosis. This approach is valid if the rates of ligand
trafficking are such that the effective diffusion coefficient
resulting from transcytosis $D(\lambda,\rho)|_{p=0}$ in
Eq. (\ref{eq:lambda_ptc_eq_full}) is much smaller for all values of
$\lambda$ and $\rho$ than the extracellular diffusion constant $D_0$.

The derivation of the continuum limit of Eq. (\ref{eq:ptc_full_disc}) is
straightforward in this situation. Only the equation for the time
development of the free ligand concentration contains a linear
diffusion term. All other quantities can be described by a coupled set
of ordinary differential equations. Together, these constitute a set
of reaction-diffusion equations.  With the notation used throughout
this article, the kinetic equations in two dimensions read
\begin{eqnarray}
\label{eq:ecs_diff_limit}
{\partial_t} l & = & D_0 \triangle l -k_{\rm
  on}a^2 lr_+ + k_{\rm  off}s_+ - e_{\rm deg} l\cr
{\partial_t} s_+ & = & k_{\rm on}a^2 lr_+ - \left(b_{\rm
  int}+k_{\rm off}\right)s_+ + b_{\rm ext} s_i\cr
{\partial_t} s_i & = & b_{\rm int}s_+-\left(b_{\rm
  ext} +b_{\rm deg}\right)s_i\cr
{\partial_t} r_+ & = & \frac{f_{\rm syn}}{a^2} + k_{\rm off} s_+ +
f_{\rm ext}r_i-k_{\rm
  on}lr_+-f_{\rm int} r_+\cr
{\partial_t} r_i & = & 
f_{\rm int} r_+-\left(f_{\rm ext}+f_{\rm deg}\right)r_i.
\end{eqnarray}
This is essentially the system that was previously studied in
\cite{land02,krus04}. This reaction-diffusion system can reproduce
some experimental results that were obtained for the morphogen Dpp and
its receptor Tkv in the wing disk \cite{land02,krus04}. However, there
are experimental observations that are in disagreement with the
solutions of Eq. (\ref{eq:ecs_diff_limit}) indicating that this is not a
correct description of the ligand and receptor kinetics in this system
\cite{krus04,entc00}.  In other experimental systems,
reaction-diffusion mechanisms may play a role in pattern formation
during development \cite{turi52,mein82,koch94}.

%\bibliography{../bib/ptc_2}

\pagebreak

\begin{figure}[t]
\begin{center}
\caption{Schematic representation of ligand transport by transcytosis
  in a chain of cells of diameter $a$ indexed by $n$. The rates of
  ligand-receptor binding and un-binding, internalization and
  externalization of ligand-receptor pairs are denoted $k_{\rm on}$,
  $k_{\rm off}$, $b_{\rm int}$ and $b_{\rm ext}$.  Degradation of
  ligand occurs inside the cells with rate $b_{\rm deg}$ and in the
  extracellular space with rate $e_{\rm deg}$. Ligands can also hop
  directly between neighboring extracellular spaces at a rate $2
  D_0/a^2$ which describes their movement in the extracellular space
  around the cells by passive diffusion with diffusion coefficient
  $D_0$ (not shown).  Figure modified from \cite{boll05}. }
\label{fig:ptc_scheme}
\end{center}
\end{figure}

\begin{figure}
\begin{center}
\caption{Time development of gradient formation in our description of
  ligand transport. Ligand densities in the presence of a source at
  $x=0$ at different times $t b_{\rm deg}=0.72, 2.16, 3.6$ during
  gradient formation (black lines) and in steady state (red
  lines). Lines indicate solutions to Eq. (\ref{eq:lambda_ptc_eq_full}),
  while symbols indicate solutions to Eq. (\ref{eq:ptc_full_disc}) for
  comparison.  (A-C) Time development of the profiles of the total
  ligand density $\lambda(x,t)$ (A), the total receptor density
  $\rho(x,t)$ (B), and the receptor bound ligand density
  $s_i(x,t)+s_+(x,t)$ (C) in the absence of extracellular diffusion,
  i.e. for $D_0=0$.  (D) Like A but with $2D_0/a^2b_{\rm deg}=10/3$,
  i.e. in the presence of extracellular diffusion. All concentrations
  are normalized to the steady state value of the surface receptor
  concentration in the absence of ligands $r_0$.  Initial conditions
  at $t=0$: $\lambda(x)=0$ and $\rho(x)=(1+f_{\rm int}/f_{\rm
  ext})r_0$.  Parameters are: $k_{\rm off}/b_{\rm deg}=b_{\rm
  int}/b_{\rm deg}=f_{\rm int}/b_{\rm deg}=1000/3$, $k_{\rm on}
  ar_{\rm max}/b_{\rm deg}=8000/3$, $b_{\rm ext}/b_{\rm deg}=f_{\rm
  ext}/b_{\rm deg}=2000/3$, $e_{\rm deg}/b_{\rm deg}=2/3$, $f_{\rm
  deg}/b_{\rm deg}=1$, $f_{\rm syn}^0/a r_{\rm max}b_{\rm deg}=1/12$,
  $\psi=2$, $j_{0}/b_{\rm deg}ar_{\rm max}=25/6$, $r_0/r_{\rm
  max}=1/7$, and $j=0$ at $x/a=50$.  }
\label{fig:comp_rec_dyn}
\end{center}
\end{figure}

\begin{figure}
\begin{center}
\caption{ Effective transport coefficients and degradation rates for
  transcytosis.  (A,B) Coefficients $D_\lambda(\lambda,\rho)$,
  $D_\rho(\lambda,\rho)$, and $k_\lambda(\lambda,\rho)$ in the
  transport equation (\ref{eq:lambda_ptc_eq_full}) as a function of
  the dimensionless ratio $\lambda/\rho$ of the total ligand
  concentration $\lambda$ and the total receptor concentration $\rho$.
  The solid lines show the coefficients $D_\lambda$ and $D_\rho$ in
  presence of extracellular diffusion with $2D_0/a^2b_{\rm deg}=10/3$.
  Broken lines show these coefficients in the absence of extracellular
  diffusion, i.e. with $D_0=0$.  Inset in A: the solid line shows
  $k_\lambda$ with $e_{\rm deg}/b_{\rm deg}=2/3$ and the broken line
  with $e_{\rm deg}=0$.  $\rho_0$ denotes the steady state total
  receptor concentration in the absence of ligands. (C) Effective
  diffusion coefficient $D(\lambda)$ in the transport equation for the
  constant surface receptor approximation (\ref{eq:dldt_rconst}) as a
  function of the ligand concentration $\lambda/r$ for $2D_0/a^2b_{\rm
  deg}=10/3$ (solid line) and $D_0=0$ (dashed line). Inset: effective
  degradation rate $k(\lambda)$ as a function of $\lambda/r$ for
  $e_{\rm deg}=0$ (dashed line) and $e_{\rm deg}/b_{\rm deg}=2/3$
  (solid line).  The constant total surface receptor concentration is
  denoted $r$.  Parameters as in Fig.~\ref{fig:comp_rec_dyn} with
  $k_{\rm on} a\rho/b_{\rm deg}=8000/3$ in A,B and $k_{\rm on}
  ar/b_{\rm deg}=8000/3$ in C. }
\label{fig:dlambda}
\end{center}
\end{figure}

\begin{figure}
\caption{ Ligand transport by transcytosis with a directional bias.
(A) Drift velocity $V_\beta$ from Eq. (\ref{eq:vbeta_expl}) as a
function of $\lambda/\rho$.  (B,C) Time development of gradient
formation with directional bias.  Profiles of the total ligand
concentration $\lambda(x,t)$ (B) and the total receptor concentration
$\rho(x,t)$ (C) in the presence of a source at $x=0$ at different
times $t b_{\rm deg}=0.72, 2.16, 3.6$ during gradient formation (black
lines) and in steady state (red lines). Lines indicate solutions to
Eq. (\ref{eq:ptc_eq_full_bias}), while symbols indicate solutions to
Eq. (\ref{eq:ptc_bias_disc}) for comparison.  All concentrations are
normalized to the steady state value of the surface receptor
concentration in the absence of ligands $r_0$.  Initial condition:
$\lambda(x)=0$ and $\rho(x)=(1+f_{\rm int}/f_{\rm ext})r_0$.
Parameters as in Fig.~\ref{fig:comp_rec_dyn} with $\beta=.1$, $D_0=0$,
and $j=0$ at $x/a=100$. }
\label{fig:bias}
\end{figure}

\begin{figure}[t]
\begin{center}
\caption{Non-robust and robust steady state gradients in our
  description of morphogen transport with constant surface receptor
  concentration given by Eq. (\ref{eq:dldt_rconst}).  (A) Ligand
  profiles in the steady state for $D_0=0$ and $j_0/b_{\rm deg}R=7$
  where robustness is small, ${\mathcal{R}}\simeq 0.1$. The profile is
  strongly affected by halving (dotted line) or doubling (dashed line)
  the ligand current of the reference state (solid line).  The
  positions of an arbitrarily chosen concentration threshold are
  indicated. (B) Ligand profiles in the steady state for $D_0=0$ and
  $j_0/b_{\rm deg}R=70$ where robustness is large,
  ${\mathcal{R}}\simeq $470. The reference profile (solid line) is
  almost unaffected by halving (dotted line) or doubling (dashed line,
  covered by the solid line) the ligand current of the reference
  state.  (C) Like B but with extracellular diffusion ($D_0/a^2b_{\rm
  deg}=50$) which reduces robustness to ${\mathcal{R}}\simeq
  0.32$. The insets in A-C show the respective profiles of the
  receptor-bound ligand concentration $s_++s_i$ which is the
  biologically relevant quantity that triggers signal transduction in
  the cells.  (D) Robustness ${\mathcal{R}}$ of steady state ligand
  profiles as a function of the ligand current $j_{0}$ from the source
  for different values of the ratio of the extracellular diffusion
  length $\xi_{d}=\sqrt{D_0/e_{\rm deg}}$ and the cell size $a$.  The
  description of morphogen transport with constant surface receptor
  approximation given by Eq. (\ref{eq:dldt_rconst}) was used to
  calculate ${\mathcal{R}}$.  Figure modified from \cite{boll05}.
  Parameters are $b_{\rm int}/b_{\rm deg}=b_{\rm ext}/b_{\rm deg}=
  3\times 10^3$, $k_{\rm on}ar/b_{\rm deg}=1.1\times 10^4$, $e_{\rm
  deg}/b_{\rm deg}=5$, and $k_{\rm off}/b_{\rm deg}=7\times 10^2$.}
\label{fig:robu_illu}
\end{center}
\end{figure} 

\begin{figure}[tp]
\begin{center}
\caption{Ligand transport in two dimensions. (A) Tissue in the region
  of the wing disk where the Dpp gradient forms. Cell membranes are
  labeled in red, the morphogen Dpp is shown in green.  (B) Schematic
  of this tissue. Source cells which produce ligands are shown in
  darker gray, the cells in the target tissue in light gray.  (C) The
  rates for the various cellular processes are denoted as in one
  dimension, see Fig.~\ref{fig:ptc_scheme}. For the triangular lattice
  with hexagonal cells which we use here, receptor bound ligands can
  be present on the six edges of each cell and inside the cells. The
  concentration of the receptor-bound ligands on edge $j$ of cell $n$
  is denoted $S_{n,j}$, that inside cell $n$ is termed $S_n^{\rm
  (i)}$.  Free ligands exist in the gaps between two cell
  surfaces. Their concentration in gap $\langle n,j\rangle$ which is
  located adjacent to edge $j$ of cell $n$ is denoted $L_{\langle
  n,j\rangle}$. (D) Triangular lattice structure with hexagonal cells
  used in our discrete theoretical description of ligand transport in
  two dimensions.  This lattice structure approximates the situation
  shown in B.  The source cells which are shown in darker gray secrete
  ligands into the extracellular spaces surrounding them. The cell
  diameter is $a$.}
\label{fig:lattices_2d}
\end{center}
\end{figure}

\begin{figure}
\begin{center}
\caption{Time development of ligand densities $\lambda(\vec x,t)$ in
  two dimensions. The solution to the continuum transport equation
  (\ref{eq:dldt_rconst_2d}) is compared to that of the discrete
  description given by Eq. (\ref{eq:disc_const_r_2d}) in presence of a
  region which the ligand cannot enter located at $6\leq x/a\leq 11$
  and $-4\leq y/a \leq 4$.  In the discrete description, this is
  realized by setting $b_{\rm int}$ to zero in the region. In the
  continuum description, zero flux boundary conditions are imposed on
  the outlining edge of the region.  (A-D) Two dimensional ligand
  profiles at $t b_{\rm deg}=3.6$ (A,B) and at $t b_{\rm deg}=17.3$
  which is close to the steady state (C,D). These profiles were
  obtained by solving the discrete (A,C) and continuum description
  (B,D) respectively. The rectangular region which the ligand cannot
  enter appears in white. It appears smaller in A and C than in B and
  D because in the discrete description, ligands can still bind to
  surface receptors on the cells located at the edge of the region so
  that $\lambda>0$ for these cells. (E-G) Profiles of $\lambda(\vec
  x,t)$ along the slices indicated in B at $t b_{\rm deg}=0.72, 2.16,
  3.6$ (black lines) and in steady state (red lines).  Lines indicate
  solutions to Eq. (\ref{eq:dldt_rconst_2d}), while symbols indicate
  solutions to Eq. (\ref{eq:disc_const_r_2d}) for comparison. (H)
  Contrast of the depletion of $\lambda(\vec x,t)$ shown in G. The
  contrast is defined as $c(t) = 1 - \lambda_b(t)/\lambda_o(t)$ where
  $\lambda_b(t)$ and $\lambda_o(t)$ are the total ligand concentration
  at the locations shown by the crosses in A, i.e. directly behind the
  clone and far away from it, respectively. Initial condition:
  $\lambda(\vec x)=0$.  Parameters as in Fig.~\ref{fig:comp_rec_dyn}
  with $D_0=0$, $j_0 /b_{\rm deg} ar=25/3$ at $x=0$, $j=0$ at $y/a=\pm
  25$ and at $x/a=50$. }
\label{fig:2dcont}
\end{center}
\end{figure}

\end{document}